\documentclass[english,aps , prl , reprint ,inputenc=utf8,superscriptaddress]{revtex4-1}
\usepackage[T1]{fontenc}
\usepackage[utf8]{inputenc}
\usepackage{color}
\usepackage{babel}
\usepackage{amsmath}
\usepackage{graphicx}
\usepackage[dvipsnames]{xcolor}
\newcommand{\HB}[1]{\textcolor{black}{ #1}}
\newcommand{\MLB}[1]{\textcolor{black}{ #1}}
\usepackage[unicode=true,pdfusetitle,
 bookmarks=true,bookmarksnumbered=false,bookmarksopen=false,
 breaklinks=false,pdfborder={0 0 1},backref=false,colorlinks=false]
 {hyperref}
\hypersetup{
 colorlinks=true,citecolor=blue,linkcolor=blue}
\begin{document}
\title{Coupled interactions at the ionic graphene/water interface}
\author{Anton Robert}
\email{anton.robert@ens.fr}

\affiliation{PASTEUR, Département de chimie, École normale supérieure, PSL University,
Sorbonne Université, CNRS, 75005 Paris, France}
\author{Hélène Berthoumieux}
\email{helene.berthoumieux@sorbonne-universite.fr}
\affiliation{Sorbonne Université, CNRS, Laboratoire de Physique Théorique de
la Matière Condensée (LPTMC, UMR 7600), F-75005 Paris, France}
\affiliation{Fachbereich Physik, Freie Universität Berlin, Arnimallee 14, Berlin, 14195, Germany}
\author{Marie-Laure Bocquet}
\email{marie-laure.bocquet@ens.fr}
\affiliation{PASTEUR, Département de chimie, École normale supérieure, PSL University,
Sorbonne Université, CNRS, 75005 Paris, France}
\begin{abstract}
We compute ionic free energy adsorption profiles at aqueous graphene interface by developing a self-consistent approach. To do so, we design a microscopic model for water and put the liquid on an equal footing with the graphene described by its electronic band structure. By evaluating progressively the electronic/dipolar coupled electrostatic interactions, we show that the coupling level including mutual graphene/water screening permits to recover remarkably the precision of extensive quantum simulations. We further derive the potential of mean force evolution of several alkali cations.
\end{abstract}
\maketitle
The peculiar properties of the water/graphene interface have been unveiled in pioneering experimental \citep{radha_molecular_2016,secchi_massive_2016,fumagalli_anomalously_2018} and theoretical \citep{wu_graphitic_2013,tocci_friction_2014,misra_insights_2017,brandenburg_physisorption_2019,monet_nonlocal_2021,kavokine_fluctuation-induced_2022} studies.  This results in particular in the  
extraordinary transport efficiency in water-filled carbon nanotubes and nanochannels \citep{faucher_critical_2019,bocquet_nanofluidics_2020,kavokine_fluids_2021}.
Moreover, the presence of charges in the wet nanometric channels leads to exotic
ionic behaviors \citep{siria_giant_2013,esfandiar_size_2017,comtet_nanoscale_2017,mouterde_molecular_2019}
that are the cornerstone of energy storage applications \citep{salanne_efficient_2016}
and blue energy harvesting \citep{siria_new_2017}. \HB{Although experimental
data \citep{mccaffrey_mechanism_2017,iamprasertkun_capacitance_2019}
regarding specific graphene-ion interactions in water are still few
in numbers, the need to overtake classical molecular dynamics (MD) approximations and 
to model them at the same level as metal/liquid interfaces
\citep{misra_ion_2021,scalfi_molecular_2021} has been acknowledged}. Beyond classical approaches, state-of-the-art
quantum calculations combined with solvation codes \citep{williams_effective_2017,zhan_specific_2019,ruggeri_multi-scale_2021}
and even fully explicit ab initio methods \citep{grosjean_versatile_2019,joly_osmotic_2021}
-- treating both the liquid and the solid at the Born-Oppenheimer
level - represent the current state-of-the-art but their computational
cost remains prohibitive for systematic investigations. On the other
hand, recent semi-classical numerical studies have described graphene using a perfect
metal \citep{son_image-charge_2021}, a Thomas-Fermi \citep{scalfi_semiclassical_2020,schlaich_electronic_2022},
and an atomistic polarizable force field \citep{misra_ion_2021,misra_uncovering_2021}
model. Nevertheless, theses studies ignore the semimetallic band structure
of graphene.
Continuum electrostatic approaches \citep{schwinger_chapter_1998,loche_breakdown_2018}
permit to evaluate the well-known attractive ``image-charge'' electrostatic
potential in a dielectric medium. Spatial correlations of both the
fluid and the metal can \emph{a priori} be included \citep{vorotyntsev_electrostatic_1980,kornyshev_nonlocal_1980,gabovich_image_2012}
to investigate microscopic effects. However, the self-consistent electrostatic
problem is not yet addressed and collective interactions
between electrons and molecules in the liquid are only partially and
phenomenologically taken into account. \\
In this work, we develop a quantum/classical field framework
to investigate electrostatic interactions at the aqueous graphene
interface. We propose a microscopic model for the nonlocal dielectric
properties of bulk and interfacial water and compute the polarization
function of graphene from a tight-binding model. We evaluate the response
function of a nanometric slab of water confined between two graphene
sheets by including gradually coupled electrostatic interactions between
the electrons of the semimetal and the water molecules.  This allows us to
derive an accurate evolution of the potential of mean force
(PMF) for a single cation solvated in the graphene channel. Finally we explore the
as derived PMF profiles of a few alkali ions.

\paragraph*{Theoretical framework}

Our framework, detailed in SI-Sec. 1, \HB{takes roots in quantum
field theory, and uses Feynman diagrammatics to derive the Green\textquoteright s
function of the interfacial system}. As predicted by quantum chemical calculations, the graphene/water
interface presents a negligible electronic corrugation \citep{tocci_friction_2014}
and is chemically inactive with no mixing of electronic states \citep{li_influence_2012,brandenburg_physisorption_2019}. We focus on building the non-local
linear response functions $\chi$ of the system, that relates the
mean induced charge density $\langle n_{\text{ind}}\rangle$ to an
external electrostatic potential $\phi_{\text{ext}}$ generated by a charge distribution $n_{\text{ext}}$. The generic
equations used to build the Green's function $w$ of the system and
therefore the mean electrostatic potential $\langle\phi_{\text{tot}}\rangle$
can be summarized as follows:
\begin{gather}
\phi_{\text{ext}}=v*n_{\text{ext}}\ \ \ \ \ \langle n_{\text{ind}}\rangle=\chi*\phi_{\text{ext}}\nonumber \\
w=v+v*\chi*v\ \ \ \ \ \langle\phi_{\text{tot}}\rangle=w*n_{\text{ext}}\label{eq:all_first_eq}, 
\end{gather}
with $v$ can denote the bare ($v = 1/4\pi \epsilon_0 x$\MLB{, with x the distance in 3D space}) or an effective Coulomb potential and $*$ the spatial
convolution.
The starting assumption to build $\chi$ is to consider that particles
are independent and to derive an \emph{non-interacting}
response function $\chi^{(0)}$. Next, $\chi^{(0)}$ is renormalized
by considering interactions at the mean-field level: independent particles
respond to the external potential plus the mean polarization potential
$\langle\phi_{\text{pol}}\rangle$ of the other similar particles.
The induced charge density thus reads $\langle n_{\text{ind}}\rangle=\chi^{(0)}*\left[\phi_{\text{ext}}+\langle\phi_{\text{pol}}\rangle\right]$
with $\langle\phi_{\text{pol}}\rangle=v_{\text{inter}}*\langle n_{\text{ind}}\rangle$
and where $v_{\text{inter}}$ is the effective interparticle potential. This
recursive equation combined with Eq. \ref{eq:all_first_eq} gives
\begin{align}
\chi & =\chi^{(0)}+\chi^{(0)}*v_{\text{inter}}*\chi.\label{eq:mean_field_eq}
\end{align}
Eq. \ref{eq:all_first_eq} and Eq. \ref{eq:mean_field_eq} sets of equations \HB{give the definition for $v_{\text{inter}}$} and are used in the following to build the response function of the water $(\chi_{\text{w}})$
and the electronic $(\chi_{\text{e}})$ part separately, but also
to build $\chi$ or $w$ of the entire interfacial system.
The interfacial system consists of a channel of nanometric height $L$
made of two graphene sheets and filled with water. 
\paragraph*{Water bulk}

\begin{figure}
\begin{centering}
\includegraphics{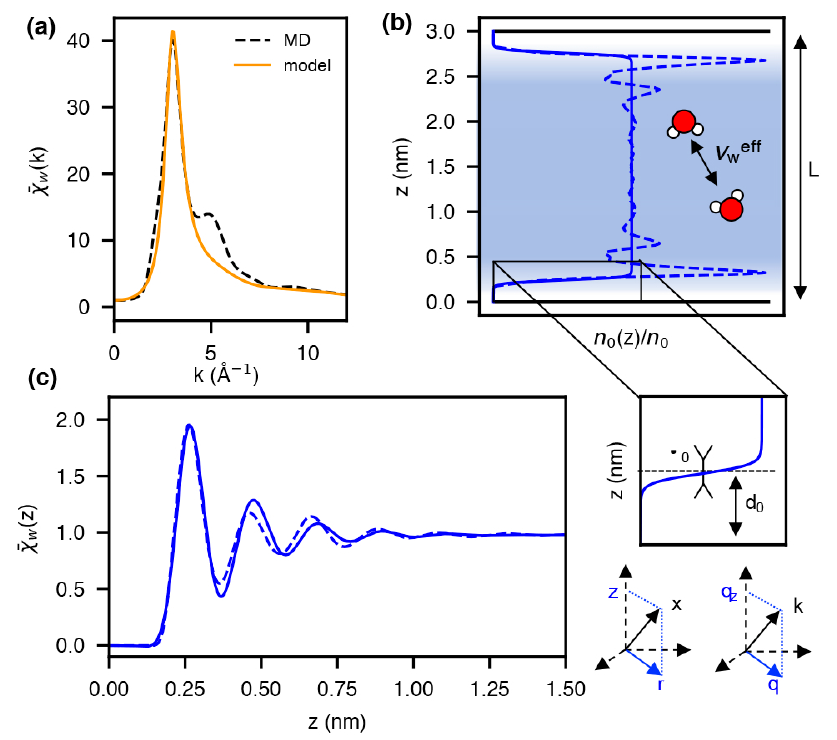}
\par\end{centering}
\caption{Dielectric response functions of water. \textbf{(a)} Susceptiblity
of SPC/E water obtained with MD \citep{jeanmairet_molecular_2016} and with the theoretical model for ($\epsilon^{\text{eff}}_{\text{w}} = 1.04$ (see SI Sec.3.1.2) , $\kappa$ = 1.65 \AA$^{-1}$, $\gamma = 0.99$) .
We show the dimensionless quantities $\bar{\chi}(k)=-\chi(k)/\epsilon_{0}k^{2}$. \textbf{(b)} Schematic drawing of the water slab and of the two considered
 molecular density profiles $n_{0}(z)$ for $L=3~\text{nm}$. The inset shows
the two parameters of the smoothed step function model for $n_{0}(z)$: $d_{0}$ and $\sigma_{0}$. \textbf{(c)}
Local dielectric susceptibility $\bar{\chi}_{\text{w}}(z)$ of the slab $P_{z}=\bar{\chi}_{\text{w}}(z)D_{z}$
corresponding to the molecular profile $n_{0}(z)$. \label{fig:wRPA}}
\end{figure}

We now build the response function $\chi_{\text{w}}$
of bulk water. Using Eq. \ref{eq:mean_field_eq},
the effective electrostatic potential in bulk water
$v_{\text{inter}}=v_{\text{w}}^{\text{eff}}$ can be written $v_{\text{w}}^{\text{eff}}(k)=1/\chi_{\text{w}}^{(0)}(k)-1/\chi(k)$,
with $k=\vert\mathbf{k}\vert$. 
The fluctuation-dissipation
theorem gives  $\chi_{\text{w}}^{(0)}(k)=-\beta S_{\text{w}}^{(0)}(k)$
where $\beta=1/k_{B}T$ and $S_{\text{w}}^{(0)}(k)$ is the single-molecule
-- or ``self'' -- charge structure factor,  and  $\chi_{\text{w}}(k)=-\beta S_{\text{w}}(k)$, $S_{\text{w}}(k)$ the charge structure factor of the liquid.
 Here, we apply this framework to the widely-used 3 point-charge model of water, SPC/E \citep{berendsen_missing_1987}. The analytical
expression of $S_{\text{w}}^{(0)}(k)$ is  given in SI-Sec.3.1.1.
$\chi(k)$ can be computed
in a MD simulation  -- e.g. the results of \citep{jeanmairet_molecular_2016} computing the polarization response function $\bar{\chi}_{\text{w}}=-\chi_{\text{w}}(k)/\epsilon_0k^2 $
that are reported in Fig. \ref{fig:wRPA}a. The sharp peak of $\bar{\chi}_{\text{w}}(k)$
centered at $k\simeq3\text{Å}$ illustrates the nonlocal and over-screening
properties of water \citep{bopp_static_1996}. \\
From the numerical
knowledge of the effective Coulomb potential for water $v_{\text{w}}^{\text{eff}}(k)$, we suggest the following
ansatz:
\begin{equation}
v_{\text{w}}^{\text{eff}}(k)=\frac{1}{\epsilon_{0}\varepsilon_{\text{w}}^{\text{eff}}}\left(\frac{1}{k^{2}}-\frac{1}{k^{2}+\kappa^{2}}-\frac{\gamma e^{-k^{2}/2\kappa^{2}}}{\kappa^{2}\sqrt{2\pi}}\right),\label{eq:v_eff_w}
\end{equation}
with the inverse screening length $\kappa$, the prefactor $\gamma$ and the effective
permittivity $\varepsilon_{\text{w}}^{\text{eff}}$ as parameters. The last one is fixed to recover the bulk dielectric permittivity of SPC/E water and can be expressed as a function of the molecular dipole moment and
bulk density of the fluid. The values of ($\kappa$, $\gamma$) are adjusted to reproduce the position and the amplitude of the over-screening
peak of $\bar{\chi}_{\text{w}}$.
The ansatz ensures $\chi_{\text{w}}(k)\rightarrow\chi_{\text{w}}^{(0)}(k)$
for $k\rightarrow\infty$  (see details in SI-Sec. 3.1.2).  \\
We plot the polarization response function derived from our framework, $ \bar{\chi}_{\text{w}}(k)=-(1/\chi_{\text{w}}^{(0)}(k)-v_{\text{w}}^{\text{eff}}(k))^{-1}/k^2\epsilon_{0}$ (orange curve, Fig.\ref{fig:wRPA}a). Our model captures nicely the dielectric properties of bulk water at low k.

\paragraph{Water slab}
We turn to the dielectric response of a water slab confined between two infinite flat interfaces in the (x,y) plane located in z=0 and z = $L$ respectively (see the sketch in Fig. \ref{fig:wRPA}b). We describe the system using cylindrical
coordinates in real and Fourier spaces, $\mathbf{x}=(\mathbf{r},z)$ with $\mathbf{r}$ lying
in the interfacial plane and $\mathbf{k=}(\mathbf{q},q_{z})$ with $q$ the in-plane Fourier
component (see Fig. \ref{fig:wRPA}c right). 
According to the in-plane invariance, the response function can be written as $\chi_{\text{w}}(q,z,z')$. We show in SI-Sec. 3.2.1 that we can
write
\begin{equation}
\chi_{\text{w}}^{(0)}(q,z,z')\simeq-\beta\frac{\sqrt{n_{0}(z)n_{0}(z')}}{n_{0}}S_{\text{w}}^{(0)}(q,\vert z-z'\vert)\label{eq:chi_0}
\end{equation}
where $S_{\text{w}}^{(0)}(q,\vert z-z'\vert)=\int\frac{\text{d}q_{z}}{2\pi}e^{iq_{z}\vert z-z'\vert}S_{\text{w}}^{(0)}(k)$
and $n_{0}(z)$ is the molecular density profile that converges to bulk
density $n_{0}$ in the middle of the channel (see Fig. \ref{fig:wRPA}b).
We assume that the water molecules interact in the slab between themselves
as in  bulk, so the slab-geometry effective potential
$v_{\text{w}}^{\text{eff}}(q,\vert z-z'\vert)$ can be obtained by Fourier transforming Eq. \ref{eq:v_eff_w} (see SI-Sec. 3.2.2). 
 To inverse Eq.\ref{eq:mean_field_eq} and carry out all subsequent computations,
we resort to matrix multiplications in the discretized space along $z$ and $z'$.
The $(i,j)^{th}$ element of the matrix $M[z_i,z_j']$ is given by the function $m(q,z_i,z_j')$.
The solution of Eq. \ref{eq:mean_field_eq} reads $X=(1-X^{(0)}V_{\text{inter}}\HB{\text({d}z)^{2})}^{-1}X^{(0)}$
where $\text{d}z=0.02 \text{Å}$ is the grid spacing and where a matrix
of size $\left\lfloor L/\text{d}z\right\rfloor ^{2}$ has been inverted.

We now derive the local dielectric susceptibility $\bar{\chi}_{\text{w}}(z)$, relating 
the response polarization field $P_{z}$ to a constant excitation
$\mathbf{D}=D_{z}\mathbf{e}_{z}$ such that $P_{z}(z)=\bar{\chi}_{\text{w}}(z)D_{z}$.
We show in SI-Sec. 4.1 that 
\begin{equation}
\bar{\chi}_{\text{w}}(z)=1-\frac{\text{d}}{\text{d}z}\left[\int_{0}^{L}\text{d}z'\varepsilon_{\text{w}}^{-1}(q\rightarrow0,z,z')z'\right],\label{eq:local_susceptibility}
\end{equation}
with  $w_{\text{w}}=\varepsilon_{\text{w}}^{-1}*v=v+v*\chi_{\text{w}}*v$,
$w_{\text{w}}$ is the Green's function of the water slab alone,
according to Eq. \ref{eq:all_first_eq}.\\
The slab \MLB{water} density profile $n_{0}(z)$ \MLB{, which describes the interaction between water and graphene}, is an input of the model (see Eq. \ref{eq:chi_0}). We first
consider a generic smoothed step function model, which captures
the vacuum layer between the fluid and a surface (encoded by $d_{0})$ and the width $(\sigma_{0})$ of the fluid
interface (inset of Fig. \ref{fig:wRPA}b). 
In agreement with previous results \citep{hansen_theory_2013,bonthuis_profile_2012,monet_nonlocal_2021},  the susceptibility calculated in this framework (solid line in Fig. \ref{fig:wRPA}c) presents an alternation of over-responding  ($\bar{\chi}_{\text{w}}(z) > \chi_b$) and under-responding ($\bar{\chi}_{\text{w}}(z) < \chi_b$) layers before reaching its bulk value $\chi_b=1-1/\varepsilon_{\text{w}}$ for z > 1.25 nm.
Refining $n_{0}(z)$ by extracting the hydrogen molecular density
from a MD simulation \citep{kavokine_fluctuation-induced_2022} (see
Fig. \ref{fig:wRPA}b) induces minor modifications in $\bar{\chi}_{\text{w}}(z)$ (dotted line in Fig. \ref{fig:wRPA}c). This first result validates our analytical microscopic model for confined water. \\

\begin{figure}
\begin{centering}
\includegraphics{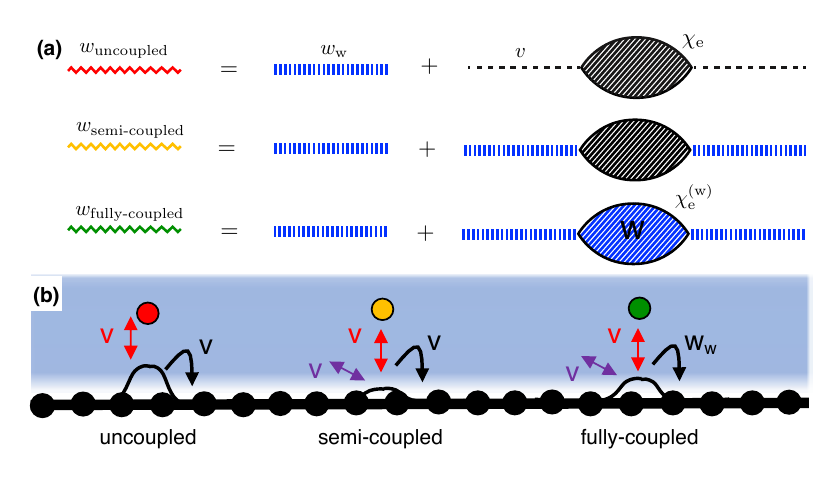}
\par\end{centering}
\caption{\textbf{(a)} Computed Feynman diagrams included in the Green's functions
for various approximations. The colors of $w$ match the one of curves
in Fig. \ref{fig:PMF}. The dashed line represents the Coulomb potential.
The hatched bubble diagram depicts $\chi_{\text{e}}$. \textbf{(b)}
Schematic illustration of the screening in the different cases (see
text for interpretation). \label{fig:diagrams}}
\end{figure}

\paragraph*{Graphene sheet}

Regarding the solid phase, the non-interacting response function $\chi_{\text{e}}^{(0)}$
can be computed and we choose a tight-binding model defined elsewhere \citep{castro_neto_electronic_2009}. Single-particle
wavefunctions $\psi_{\nu,\mathbf{p}}(\mathbf{x})$ and corresponding
eigenenergies $\epsilon_{\nu,\mathbf{p}}$ are labelled with the band
index $\nu$ and the in-plane wavevector $\mathbf{p}$. For one
graphene sheet, assuming the small spatial extent of the $p_{z}$
carbon orbitals, we consider the two-dimensional susceptibility $\chi_{\text{e}}^{(0)}(q,z,z')=\chi_{\text{e}}^{(0)}(q)\delta(z)\delta(z')$
where $\chi_{\text{e}}^{(0)}$ is (minus) the two-dimensional polarizability
given by the bare bubble diagram \citep{mahan_many-particle_1990}:

\begin{equation}
\chi_{\text{e}}^{(0)}(q)=\frac{2}{\mathcal{A}}\sum_{\nu,\mu,\mathbf{p}}\vert\lambda_{\mathbf{p},\mathbf{p+q}}^{\nu,\mu}\vert^{2}\frac{n_{F}(\epsilon_{\mu,\mathbf{p+q}})-n_{F}(\epsilon_{\nu,\mathbf{p}})}{\epsilon_{\mu,\mathbf{p}+\mathbf{q}}-\epsilon_{\nu,\mathbf{p}}},\label{eq:chi_e_0}
\end{equation}
where $\mathcal{A}$ is the surface area, $n_{F}$ the Fermi-Dirac
distribution and $\lambda_{\mathbf{p},\mathbf{p+q}}^{\nu,\mu}=\int\text{d}\mathbf{x}\psi_{\nu,\mathbf{p}}^{*}(\mathbf{x})\psi_{\mu,\mathbf{p+q}}(\mathbf{x})e^{-i\mathbf{q},\mathbf{r}}$.
We compute Eq. \ref{eq:chi_e_0} analytically at $T=0$$\text{K}$
\citep{hwang_dielectric_2007} (see result in SI-Sec. 5.1) and use
a Fermi level of $E_{F}=k_{B}T$ to include a minimal number of free
electrons. The response function $\chi_{\text{e}}$
is built from Eq. \ref{eq:mean_field_eq} using the bare Coulomb potential for the electron-electron interacting potential, $v_{\text{inter}}=v$,
which corresponds to the well-known random-phase approximation \citep{bohm_collective_1953} (see SI-Sec. 1.2). Case of two interacting graphene sheets are detailed in SI-Sec. 5.2.

\paragraph{PMF modelling and coupled interactions}

Turning to the computation of the PMF, we
first derive the  Coulomb \HB{energy at a mean field level defined as}%free energy %by thermodynamic integration
\begin{equation}
F(z)=\frac{1}{2}\iint\text{d}\mathbf{x}\text{d}\mathbf{x'}n_{\text{ext}}(\mathbf{x})\Delta w(\mathbf{x},\mathbf{x'})n_{\text{ext}}(\mathbf{x'})\label{eq:free_energy_of_solvation}
\end{equation}
where $\Delta w=w-v$. A spherical test charge of radius $b$ is placed 
in the channel at the altitude $\mathbf{x}=(0,0,z)$ such that $n_{\text{ext}}(\mathbf{x})=\pm e\delta(b-\vert\mathbf{x}-z\mathbf{e}_{z}\vert)/4\pi b^{2}$. \HB{The test charge region is assumed to respond as water}. We define the PMF as $\Delta F(z)=F(z)-F(L/2)$. \HB{It thus contains only electrostatic contributions and neglects the short-range Van der Waals interactions. }
We now gradually introduce coupled interactions in three steps labeled
uncoupled, semi-coupled and fully-coupled to build $w$ from the knowledge
of $\chi_{\text{e}}$ and $\chi_{\text{w}}$. \\
Fig. \ref{fig:diagrams}
reports the computed Feynman diagrams and the sketched coupling scenarios.
First, we consider the uncoupled case, where water and graphene
are blind to each other such that $w$ is clearly separable:
\begin{equation}
w_{\text{uncoupled}}=w_{\text{w}}+v*\chi_{\text{e}}*v.\label{eq:uncoupled}
\end{equation}
Secondly, we consider the semi-coupled scenario where the polarization charge on the graphene surface
results from the potential exerted by the ion and surrounding water
molecules. This is the sum of the bare ionic potential and the one
induced by the solvating structure of dipoles, that is the screened
potential that is obtained by the water slab Green's function $w_{\text{w}}$
and therefore
\begin{equation}
w_{\text{semi-coupled}}=w_{\text{w}}+w_{\text{w}}*\chi_{\text{e}}*w_{\text{w}}.\label{eq:semi-coupled}
\end{equation}
It is equivalent to an interfacial semi-classical
simulation adding a self-consistent optimization of the surface
polarization at each time step, taking into account fixed -- and
equal to their values in vacuum -- site-site interactions of the
atomistic model of the metal. \HB{For analytical approaches, it corresponds to the ion-metal electrostatic interaction derived in the pioneering work of Kornyshev et al.\cite{vorotyntsev_electrostatic_1980} and later \cite{kaiser_electrostatic_2017}. }
Finally, the last fully-coupled
case unveils the presence of the polar liquid for electrons of the solid. Electron-electron interactions are effectively modified
due to the presence of water, so that we introduce
the in situ response function of the metal $\chi_{\text{e}}^{(\text{w})}$
which is built from Eq. \ref{eq:mean_field_eq} with $v_{\text{inter}}=w_{\text{w}}$. \HB{Note that this coupling effect can not be included in a simple way in the standard approaches \cite{vorotyntsev_electrostatic_1980,kaiser_electrostatic_2017}}.
The most refined Green's function  systems therefore reads
\begin{equation}
w_{\text{fully-coupled}}=w_{\text{w}}+w_{\text{w}}*\chi_{\text{e}}^{(\text{w})}*w_{\text{w}}.\label{eq:fully-coupled}
\end{equation}
With the above $w$ expressions, three different PMFs can be computed using Eq. \ref{eq:free_energy_of_solvation}. Note that the double integration
of Eq. \ref{eq:free_energy_of_solvation} is made in Fourier space
and by matrix multiplication \footnote{Using $N_{\text{ext}}[z]=J_{0}(q\sqrt{b^{2}+(z-z_{0})^{2}})/2b$ with
$J_{0}$ being the zeroth order Bessel function, we compute $F(z_{0})=\frac{1}{2}\int_{0}^{+\infty}\frac{\text{d}q}{2\pi}q\left[N_{\text{ext}}^{\dagger}(W-V)N_{\text{ext}}\right](q),$
for the three Green's functions, in log-log space using $q=e^{y}E_{F}/v_{F}$,
$y\in[-1,8]$ and $N_{y}=100$ for convergence. Note that $V[z,z']=e^{-q\vert z-z'\vert}/2\epsilon_{0}q$.}. We compute all the PMFs for $L = 6~\text{nm}$ and using the first density model for $n_{0}(z)$ with $\sigma_{0} = 0.3~\text{Å}$ \citep{monet_nonlocal_2021}.The microscopic distance $d_{0}$ is determined by imposing the long-wavelength limit of the surface
charge structure factor of water at the interface \citep{kavokine_fluctuation-induced_2022}
and equals $d_{0} = 1.3\text{Å}.$ \\
 To gain insights on the electronic and water contributions to the
PMF, we decompose the free energy contribution into two terms :  
$F= F_{\text{e}} + F_{\text{w}}$, where $F_{\text{w}}$ contains
the contribution of water as in an air/water interface replacing $w$ with $w_{\text{w}}$
in Eq. \ref{eq:free_energy_of_solvation}. \MLB{We could consider other substrates by changing $\chi_{\text{e}}^{(0)}$ Eq. \ref{eq:chi_e_0}}.

\paragraph{Results \& Discussions}

\begin{figure}
\begin{centering}
\includegraphics{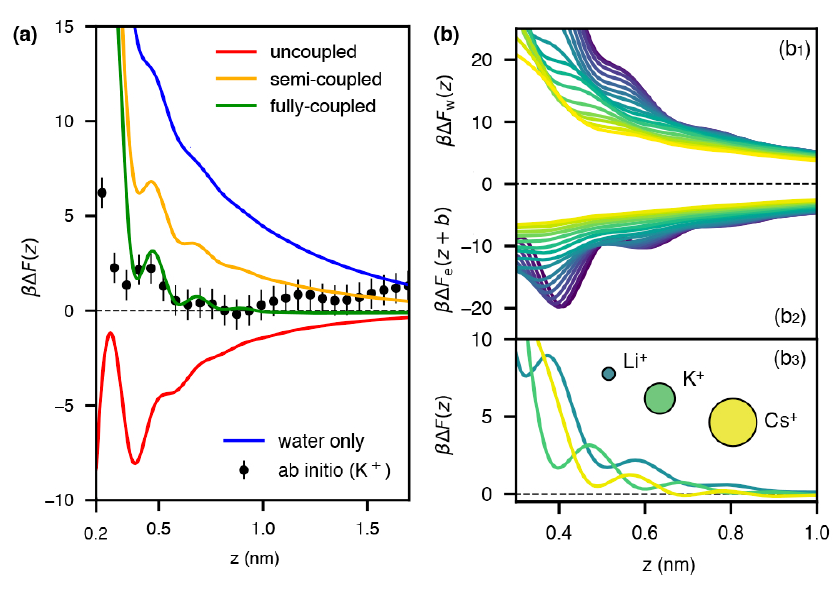}
\par\end{centering}
\caption{\textbf{(a)} PMF of $\text{K}^{+}$ ($b = 2~\text{Å}$ ) at the graphene-water interface.
Models with increasing coupling (solid lines) compared to a graphene-free model (blue line) and ab initio simulations \citep{joly_osmotic_2021} (black dots with error bars).
\textbf{(b)} Detailed contributions to the PMF from water ($b_{1})$
and from graphene $(b_{2})$ with increasing ionic radius from point charge (blue) to large radius (yellow). For $\Delta F_{\text{e}}$, the ionic center is placed at increasing altitude $\mathbf{x} = (0,0,z+b)$ for increasing radius.  $(b_{3})$ Comparative PMF for three
alkali ions. The PMF for Li$^{+}$ (resp. Cs$^{+}$)
is obtained using $b = 1~\text{Å}$ (resp. $b = 3~\text{Å}$). \label{fig:PMF}}
\end{figure}

Fig. \ref{fig:PMF}a displays the resulting different computed profiles for $\Delta F$ for one single positive charge of radius $b = 2~\text{Å}$, together with a reference curve computed recently from an ab initio\emph{ }molecular dynamics (AIMD) study, for K$^{+}$
solvated in a $2$nm thick water slab on graphene \citep{joly_osmotic_2021}. \HB{In the AIMD simulation the limited thickness of the water slab induces a second water/air interface explaining the non-monotonic and repulsive ab initio PMF behavior above 1nm. This large range of graphene-water distance (1$< z <$2 nm) is not meaningful here. Hence} for sake of comparison, we shift the ab initio PMF such that it is vanishing in the middle of the water slab, for $z$ around $1$ nm (black dots, Fig. \ref{fig:PMF}a). The water contribution $\Delta F_{\text{w}}$ shows the expected repulsive behaviour of the ion at an air/interface (blue curve, Fig. \ref{fig:PMF}a).
Concerning the water/graphene interface, the uncoupled PMF profile (red curve, Fig. \ref{fig:PMF}a) is strongly attractive and presents oscillations with small amplitudes near
the surface stemming from the non-local dielectric response of water. Moreover it deviates a lot from the ab initio plot. 
 \\
 Moving to the semi-coupled PMF profile (orange curve, Fig. \ref{fig:PMF}a), its energy position is shifted to positive values fingerprinting a long-range repulsion and a net reduction of the graphene-ion interaction due to surrounding
water molecules. Interestingly this result is in quantitative agreement with semi-classical simulations \citep{misra_ion_2021,scalfi_microscopic_2021,son_image-charge_2021}
using \emph{ad hoc} surface polarization models (SI-Section 6.2). \\
Finally the fully-coupled PMF curve (green curve in Fig. \ref{fig:PMF}a) reveals
a re-amplification of the wall-ion attraction by several thermal energy
units and matches almost quantitatively the ab initio PMF. This is the key finding of our approach. The nice agreement suggests that this semi-analytical approach incorporating electrostatics in a self-consistent way, is
able to reproduce some key features of the state-of-art reference PMF like the position
and amplitude of the three local minima. The stabilizing effect present in the fully-coupled case can be qualitatively understood as follows. 
The absence of repulsive interaction between charge carriers
would make them accumulate to one point in order to screen the ionic potential.
Thanks to electron-electron interactions a finite polarization charge
can accumulate on the surface as shown by the uncoupled case cartoon
in Fig. \ref{fig:diagrams}b. Water molecules actually screen the
ionic potential and reduce the polarization charge (semi-coupled case
Fig. \ref{fig:diagrams}b), but in the last fully-coupled case, the
presence of water effectively reduce electron-electron interactions
- by roughly a factor of $(\varepsilon_{\text{w}}+1)/2$ for electrons
that are far apart as shown in SI-Sec. 6.3. As a result, the polarization
charge gets re-amplified and so does the surface-ion screened potential. \HB{The plots differ significantly at short distance, $z<$0.5 nm, where non-electrostatic contributions of the PFM -not considered here - are dominant \cite{loche_breakdown_2018}. }\\
We now investigate the variations of the PMF with
varying radius $b$ ranging from point charge to $3 ~\text{Å}$ with detailed contributions from water and graphene. Fig. \ref{fig:PMF}$\text{b}_{1}$ shows that
water repels more strongly smaller ions from the interface. This can
be understood by considering the hydrated radius of the cations - \HB{defined in continuous theories as the range on which the ion polarizes the surrounding fluid} -that
is inversely proportional to the ionic radius \citep{marcus_effect_2009}.
Coming from the bulk, $\text{Li}{}^{+}$ is the first to break its
solvation shell. Fig. \ref{fig:PMF}$\text{b}_{2}$ compares the non-monotonic
surface contribution $\Delta F_{\text{e}}$ for the series of ions, which center is shifted so that
the available space for water molecules between ion and surface is equal for each ion.
We link the increasing attraction for smaller radii to the ordering
degree of the hydration shells as follows. In the limit of poorly
structured hydration shells -- e.g. for Cs$^{+}$-- we find the
monotonic surface-ion potential of an attenuated charge in vacuum.
The opposite limit is a point charge with three highly ordered hydration
shells. This gives rise to three special places where ice-like water,
with a low permittivity, is practically transparent to the potential
stemming from the polarization charge on the graphene surface. Summing
both contributions in Fig. \ref{fig:PMF}$\text{b}_{3}$ for three
cations in the alkali series leads to complex PMF profiles. We observe that for
increasing radius the three local minima are stabilized \HB{in energy in agreement with an increased capacitance \citep{iamprasertkun_capacitance_2019} and a reduced hydration energy \citep{zhan_specific_2019}. Indeed small ions like $\text{Li}{}^{+}$ manifest a strong solvation environment difficult to break hampering its adsorption. Proceeding down the series, $\text{Cs}{}^{+}$ yields a weak solvation shell which can be easily desolvated at the graphene interface.}

\paragraph{Conclusion}
In this letter, we build a self-consistent theoretical framework which permits to investigate analytically the single ionic adsorption at the graphene/water interface. 
By including the semimetallic band structure of graphene, building a microscopic model for interfacial water and considering the mutual screening of the two materials, we obtained
results that are in excellent agreement with expensive quantum free
energy perturbation methods, at a negligible computational cost. Our PMF predictions for the alkali series are in agreement with experimental
observations and permit to distinguish the liquid water and graphene surface contributions. We hope that this versatile and generalizable method, will renew some interest in
semi-analytical approaches and be used to investigate more complex systems involving for example ion-ion interactions in nanochannels. 

\paragraph*{Acknowledgments}
A.R. thanks D. Borgis for discussions. A.R and M.-L.B. acknowledge funding from
EU H2020 Framework Programme/ERC Advanced Grant agreement number 785911-Shadoks. 
H. B. acknowledges funding from Humboldt Research Fellowship Programme for Experienced Researchers.
\section*{Code availability}
Our code is freely available in the GitHub repository \url{https://github.com/anton-smirnov-robert/pmf_water_graphene}.
\bibliographystyle{apsrev4-1}
%\bibliography{main}

\begin{thebibliography}{50}%
\makeatletter
\providecommand \@ifxundefined [1]{%
 \@ifx{#1\undefined}
}%
\providecommand \@ifnum [1]{%
 \ifnum #1\expandafter \@firstoftwo
 \else \expandafter \@secondoftwo
 \fi
}%
\providecommand \@ifx [1]{%
 \ifx #1\expandafter \@firstoftwo
 \else \expandafter \@secondoftwo
 \fi
}%
\providecommand \natexlab [1]{#1}%
\providecommand \enquote  [1]{``#1''}%
\providecommand \bibnamefont  [1]{#1}%
\providecommand \bibfnamefont [1]{#1}%
\providecommand \citenamefont [1]{#1}%
\providecommand \href@noop [0]{\@secondoftwo}%
\providecommand \href [0]{\begingroup \@sanitize@url \@href}%
\providecommand \@href[1]{\@@startlink{#1}\@@href}%
\providecommand \@@href[1]{\endgroup#1\@@endlink}%
\providecommand \@sanitize@url [0]{\catcode `\\12\catcode `\$12\catcode
  `\&12\catcode `\#12\catcode `\^12\catcode `\_12\catcode `\%12\relax}%
\providecommand \@@startlink[1]{}%
\providecommand \@@endlink[0]{}%
\providecommand \url  [0]{\begingroup\@sanitize@url \@url }%
\providecommand \@url [1]{\endgroup\@href {#1}{\urlprefix }}%
\providecommand \urlprefix  [0]{URL }%
\providecommand \Eprint [0]{\href }%
\providecommand \doibase [0]{http://dx.doi.org/}%
\providecommand \selectlanguage [0]{\@gobble}%
\providecommand \bibinfo  [0]{\@secondoftwo}%
\providecommand \bibfield  [0]{\@secondoftwo}%
\providecommand \translation [1]{[#1]}%
\providecommand \BibitemOpen [0]{}%
\providecommand \bibitemStop [0]{}%
\providecommand \bibitemNoStop [0]{.\EOS\space}%
\providecommand \EOS [0]{\spacefactor3000\relax}%
\providecommand \BibitemShut  [1]{\csname bibitem#1\endcsname}%
\let\auto@bib@innerbib\@empty
%</preamble>
\bibitem [{\citenamefont {Radha}\ \emph {et~al.}(2016)\citenamefont {Radha},
  \citenamefont {Esfandiar}, \citenamefont {Wang}, \citenamefont {Rooney},
  \citenamefont {Gopinadhan}, \citenamefont {Keerthi}, \citenamefont
  {Mishchenko}, \citenamefont {Janardanan}, \citenamefont {Blake},
  \citenamefont {Fumagalli}, \citenamefont {Lozada-Hidalgo}, \citenamefont
  {Garaj}, \citenamefont {Haigh}, \citenamefont {Grigorieva}, \citenamefont
  {Wu},\ and\ \citenamefont {Geim}}]{radha_molecular_2016}%
  \BibitemOpen
  \bibfield  {author} {\bibinfo {author} {\bibfnamefont {B.}~\bibnamefont
  {Radha}}, \bibinfo {author} {\bibfnamefont {A.}~\bibnamefont {Esfandiar}},
  \bibinfo {author} {\bibfnamefont {F.~C.}\ \bibnamefont {Wang}}, \bibinfo
  {author} {\bibfnamefont {A.~P.}\ \bibnamefont {Rooney}}, \bibinfo {author}
  {\bibfnamefont {K.}~\bibnamefont {Gopinadhan}}, \bibinfo {author}
  {\bibfnamefont {A.}~\bibnamefont {Keerthi}}, \bibinfo {author} {\bibfnamefont
  {A.}~\bibnamefont {Mishchenko}}, \bibinfo {author} {\bibfnamefont
  {A.}~\bibnamefont {Janardanan}}, \bibinfo {author} {\bibfnamefont
  {P.}~\bibnamefont {Blake}}, \bibinfo {author} {\bibfnamefont
  {L.}~\bibnamefont {Fumagalli}}, \bibinfo {author} {\bibfnamefont
  {M.}~\bibnamefont {Lozada-Hidalgo}}, \bibinfo {author} {\bibfnamefont
  {S.}~\bibnamefont {Garaj}}, \bibinfo {author} {\bibfnamefont {S.~J.}\
  \bibnamefont {Haigh}}, \bibinfo {author} {\bibfnamefont {I.~V.}\ \bibnamefont
  {Grigorieva}}, \bibinfo {author} {\bibfnamefont {H.~A.}\ \bibnamefont {Wu}},
  \ and\ \bibinfo {author} {\bibfnamefont {A.~K.}\ \bibnamefont {Geim}},\
  }\href {\doibase 10.1038/nature19363} {\bibfield  {journal} {\bibinfo
  {journal} {Nature}\ }\textbf {\bibinfo {volume} {538}},\ \bibinfo {pages}
  {222} (\bibinfo {year} {2016})}\BibitemShut {NoStop}%
\bibitem [{\citenamefont {Secchi}\ \emph {et~al.}(2016)\citenamefont {Secchi},
  \citenamefont {Marbach}, \citenamefont {Nigu{\`e}s}, \citenamefont {Stein},
  \citenamefont {Siria},\ and\ \citenamefont {Bocquet}}]{secchi_massive_2016}%
  \BibitemOpen
  \bibfield  {author} {\bibinfo {author} {\bibfnamefont {E.}~\bibnamefont
  {Secchi}}, \bibinfo {author} {\bibfnamefont {S.}~\bibnamefont {Marbach}},
  \bibinfo {author} {\bibfnamefont {A.}~\bibnamefont {Nigu{\`e}s}}, \bibinfo
  {author} {\bibfnamefont {D.}~\bibnamefont {Stein}}, \bibinfo {author}
  {\bibfnamefont {A.}~\bibnamefont {Siria}}, \ and\ \bibinfo {author}
  {\bibfnamefont {L.}~\bibnamefont {Bocquet}},\ }\href {\doibase
  10.1038/nature19315} {\bibfield  {journal} {\bibinfo  {journal} {Nature}\
  }\textbf {\bibinfo {volume} {537}},\ \bibinfo {pages} {210} (\bibinfo {year}
  {2016})}\BibitemShut {NoStop}%
\bibitem [{\citenamefont {Fumagalli}\ \emph {et~al.}(2018)\citenamefont
  {Fumagalli}, \citenamefont {Esfandiar}, \citenamefont {Fabregas},
  \citenamefont {Hu}, \citenamefont {Ares}, \citenamefont {Janardanan},
  \citenamefont {Yang}, \citenamefont {Radha}, \citenamefont {Taniguchi},
  \citenamefont {Watanabe}, \citenamefont {Gomila}, \citenamefont {Novoselov},\
  and\ \citenamefont {Geim}}]{fumagalli_anomalously_2018}%
  \BibitemOpen
  \bibfield  {author} {\bibinfo {author} {\bibfnamefont {L.}~\bibnamefont
  {Fumagalli}}, \bibinfo {author} {\bibfnamefont {A.}~\bibnamefont
  {Esfandiar}}, \bibinfo {author} {\bibfnamefont {R.}~\bibnamefont {Fabregas}},
  \bibinfo {author} {\bibfnamefont {S.}~\bibnamefont {Hu}}, \bibinfo {author}
  {\bibfnamefont {P.}~\bibnamefont {Ares}}, \bibinfo {author} {\bibfnamefont
  {A.}~\bibnamefont {Janardanan}}, \bibinfo {author} {\bibfnamefont
  {Q.}~\bibnamefont {Yang}}, \bibinfo {author} {\bibfnamefont {B.}~\bibnamefont
  {Radha}}, \bibinfo {author} {\bibfnamefont {T.}~\bibnamefont {Taniguchi}},
  \bibinfo {author} {\bibfnamefont {K.}~\bibnamefont {Watanabe}}, \bibinfo
  {author} {\bibfnamefont {G.}~\bibnamefont {Gomila}}, \bibinfo {author}
  {\bibfnamefont {K.~S.}\ \bibnamefont {Novoselov}}, \ and\ \bibinfo {author}
  {\bibfnamefont {A.~K.}\ \bibnamefont {Geim}},\ }\href {\doibase
  10.1126/science.aat4191} {\bibfield  {journal} {\bibinfo  {journal}
  {Science}\ }\textbf {\bibinfo {volume} {360}},\ \bibinfo {pages} {1339}
  (\bibinfo {year} {2018})}\BibitemShut {NoStop}%
\bibitem [{\citenamefont {Wu}\ and\ \citenamefont
  {Aluru}(2013)}]{wu_graphitic_2013}%
  \BibitemOpen
  \bibfield  {author} {\bibinfo {author} {\bibfnamefont {Y.}~\bibnamefont
  {Wu}}\ and\ \bibinfo {author} {\bibfnamefont {N.~R.}\ \bibnamefont {Aluru}},\
  }\href {\doibase 10.1021/jp402051t} {\bibfield  {journal} {\bibinfo
  {journal} {The Journal of Physical Chemistry B}\ }\textbf {\bibinfo {volume}
  {117}},\ \bibinfo {pages} {8802} (\bibinfo {year} {2013})}\BibitemShut
  {NoStop}%
\bibitem [{\citenamefont {Tocci}\ \emph {et~al.}(2014)\citenamefont {Tocci},
  \citenamefont {Joly},\ and\ \citenamefont
  {Michaelides}}]{tocci_friction_2014}%
  \BibitemOpen
  \bibfield  {author} {\bibinfo {author} {\bibfnamefont {G.}~\bibnamefont
  {Tocci}}, \bibinfo {author} {\bibfnamefont {L.}~\bibnamefont {Joly}}, \ and\
  \bibinfo {author} {\bibfnamefont {A.}~\bibnamefont {Michaelides}},\ }\href
  {\doibase 10.1021/nl502837d} {\bibfield  {journal} {\bibinfo  {journal} {Nano
  Letters}\ }\textbf {\bibinfo {volume} {14}},\ \bibinfo {pages} {6872}
  (\bibinfo {year} {2014})}\BibitemShut {NoStop}%
\bibitem [{\citenamefont {Misra}\ and\ \citenamefont
  {Blankschtein}(2017)}]{misra_insights_2017}%
  \BibitemOpen
  \bibfield  {author} {\bibinfo {author} {\bibfnamefont {R.~P.}\ \bibnamefont
  {Misra}}\ and\ \bibinfo {author} {\bibfnamefont {D.}~\bibnamefont
  {Blankschtein}},\ }\href {\doibase 10.1021/acs.jpcc.7b08891} {\bibfield
  {journal} {\bibinfo  {journal} {The Journal of Physical Chemistry C}\
  }\textbf {\bibinfo {volume} {121}},\ \bibinfo {pages} {28166} (\bibinfo
  {year} {2017})}\BibitemShut {NoStop}%
\bibitem [{\citenamefont {Brandenburg}\ \emph {et~al.}(2019)\citenamefont
  {Brandenburg}, \citenamefont {Zen}, \citenamefont {Fitzner}, \citenamefont
  {Ramberger}, \citenamefont {Kresse}, \citenamefont {Tsatsoulis},
  \citenamefont {Gr{\"u}neis}, \citenamefont {Michaelides},\ and\ \citenamefont
  {Alf{\`e}}}]{brandenburg_physisorption_2019}%
  \BibitemOpen
  \bibfield  {author} {\bibinfo {author} {\bibfnamefont {J.~G.}\ \bibnamefont
  {Brandenburg}}, \bibinfo {author} {\bibfnamefont {A.}~\bibnamefont {Zen}},
  \bibinfo {author} {\bibfnamefont {M.}~\bibnamefont {Fitzner}}, \bibinfo
  {author} {\bibfnamefont {B.}~\bibnamefont {Ramberger}}, \bibinfo {author}
  {\bibfnamefont {G.}~\bibnamefont {Kresse}}, \bibinfo {author} {\bibfnamefont
  {T.}~\bibnamefont {Tsatsoulis}}, \bibinfo {author} {\bibfnamefont
  {A.}~\bibnamefont {Gr{\"u}neis}}, \bibinfo {author} {\bibfnamefont
  {A.}~\bibnamefont {Michaelides}}, \ and\ \bibinfo {author} {\bibfnamefont
  {D.}~\bibnamefont {Alf{\`e}}},\ }\href {\doibase 10.1021/acs.jpclett.8b03679}
  {\bibfield  {journal} {\bibinfo  {journal} {The Journal of Physical Chemistry
  Letters}\ ,\ \bibinfo {pages} {358}} (\bibinfo {year} {2019})}\BibitemShut
  {NoStop}%
\bibitem [{\citenamefont {Monet}\ \emph {et~al.}(2021)\citenamefont {Monet},
  \citenamefont {Bresme}, \citenamefont {Kornyshev},\ and\ \citenamefont
  {Berthoumieux}}]{monet_nonlocal_2021}%
  \BibitemOpen
  \bibfield  {author} {\bibinfo {author} {\bibfnamefont {G.}~\bibnamefont
  {Monet}}, \bibinfo {author} {\bibfnamefont {F.}~\bibnamefont {Bresme}},
  \bibinfo {author} {\bibfnamefont {A.}~\bibnamefont {Kornyshev}}, \ and\
  \bibinfo {author} {\bibfnamefont {H.}~\bibnamefont {Berthoumieux}},\ }\href
  {\doibase 10.1103/PhysRevLett.126.216001} {\bibfield  {journal} {\bibinfo
  {journal} {Physical Review Letters}\ }\textbf {\bibinfo {volume} {126}},\
  \bibinfo {pages} {216001} (\bibinfo {year} {2021})}\BibitemShut {NoStop}%
\bibitem [{\citenamefont {Kavokine}\ \emph {et~al.}(2022)\citenamefont
  {Kavokine}, \citenamefont {Bocquet},\ and\ \citenamefont
  {Bocquet}}]{kavokine_fluctuation-induced_2022}%
  \BibitemOpen
  \bibfield  {author} {\bibinfo {author} {\bibfnamefont {N.}~\bibnamefont
  {Kavokine}}, \bibinfo {author} {\bibfnamefont {M.-L.}\ \bibnamefont
  {Bocquet}}, \ and\ \bibinfo {author} {\bibfnamefont {L.}~\bibnamefont
  {Bocquet}},\ }\href {https://www.nature.com/articles/s41586-021-04284-7}
  {\bibfield  {journal} {\bibinfo  {journal} {Nature}\ }\textbf {\bibinfo
  {volume} {602}},\ \bibinfo {pages} {84} (\bibinfo {year} {2022})}\BibitemShut
  {NoStop}%
\bibitem [{\citenamefont {Faucher}\ \emph {et~al.}(2019)\citenamefont
  {Faucher}, \citenamefont {Aluru}, \citenamefont {Bazant}, \citenamefont
  {Blankschtein}, \citenamefont {Brozena}, \citenamefont {Cumings},
  \citenamefont {Pedro~de Souza}, \citenamefont {Elimelech}, \citenamefont
  {Epsztein}, \citenamefont {Fourkas}, \citenamefont {Rajan}, \citenamefont
  {Kulik}, \citenamefont {Levy}, \citenamefont {Majumdar}, \citenamefont
  {Martin}, \citenamefont {McEldrew}, \citenamefont {Misra}, \citenamefont
  {Noy}, \citenamefont {Pham}, \citenamefont {Reed}, \citenamefont {Schwegler},
  \citenamefont {Siwy}, \citenamefont {Wang},\ and\ \citenamefont
  {Strano}}]{faucher_critical_2019}%
  \BibitemOpen
  \bibfield  {author} {\bibinfo {author} {\bibfnamefont {S.}~\bibnamefont
  {Faucher}}, \bibinfo {author} {\bibfnamefont {N.}~\bibnamefont {Aluru}},
  \bibinfo {author} {\bibfnamefont {M.~Z.}\ \bibnamefont {Bazant}}, \bibinfo
  {author} {\bibfnamefont {D.}~\bibnamefont {Blankschtein}}, \bibinfo {author}
  {\bibfnamefont {A.~H.}\ \bibnamefont {Brozena}}, \bibinfo {author}
  {\bibfnamefont {J.}~\bibnamefont {Cumings}}, \bibinfo {author} {\bibfnamefont
  {J.}~\bibnamefont {Pedro~de Souza}}, \bibinfo {author} {\bibfnamefont
  {M.}~\bibnamefont {Elimelech}}, \bibinfo {author} {\bibfnamefont
  {R.}~\bibnamefont {Epsztein}}, \bibinfo {author} {\bibfnamefont {J.~T.}\
  \bibnamefont {Fourkas}}, \bibinfo {author} {\bibfnamefont {A.~G.}\
  \bibnamefont {Rajan}}, \bibinfo {author} {\bibfnamefont {H.~J.}\ \bibnamefont
  {Kulik}}, \bibinfo {author} {\bibfnamefont {A.}~\bibnamefont {Levy}},
  \bibinfo {author} {\bibfnamefont {A.}~\bibnamefont {Majumdar}}, \bibinfo
  {author} {\bibfnamefont {C.}~\bibnamefont {Martin}}, \bibinfo {author}
  {\bibfnamefont {M.}~\bibnamefont {McEldrew}}, \bibinfo {author}
  {\bibfnamefont {R.~P.}\ \bibnamefont {Misra}}, \bibinfo {author}
  {\bibfnamefont {A.}~\bibnamefont {Noy}}, \bibinfo {author} {\bibfnamefont
  {T.~A.}\ \bibnamefont {Pham}}, \bibinfo {author} {\bibfnamefont
  {M.}~\bibnamefont {Reed}}, \bibinfo {author} {\bibfnamefont {E.}~\bibnamefont
  {Schwegler}}, \bibinfo {author} {\bibfnamefont {Z.}~\bibnamefont {Siwy}},
  \bibinfo {author} {\bibfnamefont {Y.}~\bibnamefont {Wang}}, \ and\ \bibinfo
  {author} {\bibfnamefont {M.}~\bibnamefont {Strano}},\ }\href
  {https://doi.org/10.1021/acs.jpcc.9b02178} {\bibfield  {journal} {\bibinfo
  {journal} {The Journal of Physical Chemistry C}\ }\textbf {\bibinfo {volume}
  {123}},\ \bibinfo {pages} {21309} (\bibinfo {year} {2019})}\BibitemShut
  {NoStop}%
\bibitem [{\citenamefont {Bocquet}(2020)}]{bocquet_nanofluidics_2020}%
  \BibitemOpen
  \bibfield  {author} {\bibinfo {author} {\bibfnamefont {L.}~\bibnamefont
  {Bocquet}},\ }\href {\doibase 10.1038/s41563-020-0625-8} {\bibfield
  {journal} {\bibinfo  {journal} {Nature Materials}\ }\textbf {\bibinfo
  {volume} {19}},\ \bibinfo {pages} {254} (\bibinfo {year} {2020})}\BibitemShut
  {NoStop}%
\bibitem [{\citenamefont {Kavokine}\ \emph {et~al.}(2021)\citenamefont
  {Kavokine}, \citenamefont {Netz},\ and\ \citenamefont
  {Bocquet}}]{kavokine_fluids_2021}%
  \BibitemOpen
  \bibfield  {author} {\bibinfo {author} {\bibfnamefont {N.}~\bibnamefont
  {Kavokine}}, \bibinfo {author} {\bibfnamefont {R.~R.}\ \bibnamefont {Netz}},
  \ and\ \bibinfo {author} {\bibfnamefont {L.}~\bibnamefont {Bocquet}},\ }\href
  {\doibase 10.1146/annurev-fluid-071320-095958} {\bibfield  {journal}
  {\bibinfo  {journal} {Annual Review of Fluid Mechanics}\ }\textbf {\bibinfo
  {volume} {53}},\ \bibinfo {pages} {377} (\bibinfo {year} {2021})}\BibitemShut
  {NoStop}%
\bibitem [{\citenamefont {Siria}\ \emph {et~al.}(2013)\citenamefont {Siria},
  \citenamefont {Poncharal}, \citenamefont {Biance}, \citenamefont {Fulcrand},
  \citenamefont {Blase}, \citenamefont {Purcell},\ and\ \citenamefont
  {Bocquet}}]{siria_giant_2013}%
  \BibitemOpen
  \bibfield  {author} {\bibinfo {author} {\bibfnamefont {A.}~\bibnamefont
  {Siria}}, \bibinfo {author} {\bibfnamefont {P.}~\bibnamefont {Poncharal}},
  \bibinfo {author} {\bibfnamefont {A.-L.}\ \bibnamefont {Biance}}, \bibinfo
  {author} {\bibfnamefont {R.}~\bibnamefont {Fulcrand}}, \bibinfo {author}
  {\bibfnamefont {X.}~\bibnamefont {Blase}}, \bibinfo {author} {\bibfnamefont
  {S.~T.}\ \bibnamefont {Purcell}}, \ and\ \bibinfo {author} {\bibfnamefont
  {L.}~\bibnamefont {Bocquet}},\ }\href {\doibase 10.1038/nature11876}
  {\bibfield  {journal} {\bibinfo  {journal} {Nature}\ }\textbf {\bibinfo
  {volume} {494}},\ \bibinfo {pages} {455} (\bibinfo {year}
  {2013})}\BibitemShut {NoStop}%
\bibitem [{\citenamefont {Esfandiar}\ \emph {et~al.}(2017)\citenamefont
  {Esfandiar}, \citenamefont {Radha}, \citenamefont {Wang}, \citenamefont
  {Yang}, \citenamefont {Hu}, \citenamefont {Garaj}, \citenamefont {Nair},
  \citenamefont {Geim},\ and\ \citenamefont
  {Gopinadhan}}]{esfandiar_size_2017}%
  \BibitemOpen
  \bibfield  {author} {\bibinfo {author} {\bibfnamefont {A.}~\bibnamefont
  {Esfandiar}}, \bibinfo {author} {\bibfnamefont {B.}~\bibnamefont {Radha}},
  \bibinfo {author} {\bibfnamefont {F.~C.}\ \bibnamefont {Wang}}, \bibinfo
  {author} {\bibfnamefont {Q.}~\bibnamefont {Yang}}, \bibinfo {author}
  {\bibfnamefont {S.}~\bibnamefont {Hu}}, \bibinfo {author} {\bibfnamefont
  {S.}~\bibnamefont {Garaj}}, \bibinfo {author} {\bibfnamefont {R.~R.}\
  \bibnamefont {Nair}}, \bibinfo {author} {\bibfnamefont {A.~K.}\ \bibnamefont
  {Geim}}, \ and\ \bibinfo {author} {\bibfnamefont {K.}~\bibnamefont
  {Gopinadhan}},\ }\href {\doibase 10.1126/science.aan5275} {\bibfield
  {journal} {\bibinfo  {journal} {Science}\ }\textbf {\bibinfo {volume}
  {358}},\ \bibinfo {pages} {511} (\bibinfo {year} {2017})}\BibitemShut
  {NoStop}%
\bibitem [{\citenamefont {Comtet}\ \emph {et~al.}(2017)\citenamefont {Comtet},
  \citenamefont {Nigu{\`e}s}, \citenamefont {Kaiser}, \citenamefont {Coasne},
  \citenamefont {Bocquet},\ and\ \citenamefont
  {Siria}}]{comtet_nanoscale_2017}%
  \BibitemOpen
  \bibfield  {author} {\bibinfo {author} {\bibfnamefont {J.}~\bibnamefont
  {Comtet}}, \bibinfo {author} {\bibfnamefont {A.}~\bibnamefont {Nigu{\`e}s}},
  \bibinfo {author} {\bibfnamefont {V.}~\bibnamefont {Kaiser}}, \bibinfo
  {author} {\bibfnamefont {B.}~\bibnamefont {Coasne}}, \bibinfo {author}
  {\bibfnamefont {L.}~\bibnamefont {Bocquet}}, \ and\ \bibinfo {author}
  {\bibfnamefont {A.}~\bibnamefont {Siria}},\ }\href
  {http://www.nature.com/articles/nmat4880} {\bibfield  {journal} {\bibinfo
  {journal} {Nature Materials}\ }\textbf {\bibinfo {volume} {16}},\ \bibinfo
  {pages} {634} (\bibinfo {year} {2017})}\BibitemShut {NoStop}%
\bibitem [{\citenamefont {Mouterde}\ \emph {et~al.}(2019)\citenamefont
  {Mouterde}, \citenamefont {Keerthi}, \citenamefont {Poggioli}, \citenamefont
  {Dar}, \citenamefont {Siria}, \citenamefont {Geim}, \citenamefont {Bocquet},\
  and\ \citenamefont {Radha}}]{mouterde_molecular_2019}%
  \BibitemOpen
  \bibfield  {author} {\bibinfo {author} {\bibfnamefont {T.}~\bibnamefont
  {Mouterde}}, \bibinfo {author} {\bibfnamefont {A.}~\bibnamefont {Keerthi}},
  \bibinfo {author} {\bibfnamefont {A.~R.}\ \bibnamefont {Poggioli}}, \bibinfo
  {author} {\bibfnamefont {S.~A.}\ \bibnamefont {Dar}}, \bibinfo {author}
  {\bibfnamefont {A.}~\bibnamefont {Siria}}, \bibinfo {author} {\bibfnamefont
  {A.~K.}\ \bibnamefont {Geim}}, \bibinfo {author} {\bibfnamefont
  {L.}~\bibnamefont {Bocquet}}, \ and\ \bibinfo {author} {\bibfnamefont
  {B.}~\bibnamefont {Radha}},\ }\href {\doibase 10.1038/s41586-019-0961-5}
  {\bibfield  {journal} {\bibinfo  {journal} {Nature}\ }\textbf {\bibinfo
  {volume} {567}},\ \bibinfo {pages} {87} (\bibinfo {year} {2019})}\BibitemShut
  {NoStop}%
\bibitem [{\citenamefont {Salanne}\ \emph {et~al.}(2016)\citenamefont
  {Salanne}, \citenamefont {Rotenberg}, \citenamefont {Naoi}, \citenamefont
  {Kaneko}, \citenamefont {Taberna}, \citenamefont {Grey}, \citenamefont
  {Dunn},\ and\ \citenamefont {Simon}}]{salanne_efficient_2016}%
  \BibitemOpen
  \bibfield  {author} {\bibinfo {author} {\bibfnamefont {M.}~\bibnamefont
  {Salanne}}, \bibinfo {author} {\bibfnamefont {B.}~\bibnamefont {Rotenberg}},
  \bibinfo {author} {\bibfnamefont {K.}~\bibnamefont {Naoi}}, \bibinfo {author}
  {\bibfnamefont {K.}~\bibnamefont {Kaneko}}, \bibinfo {author} {\bibfnamefont
  {P.-L.}\ \bibnamefont {Taberna}}, \bibinfo {author} {\bibfnamefont {C.~P.}\
  \bibnamefont {Grey}}, \bibinfo {author} {\bibfnamefont {B.}~\bibnamefont
  {Dunn}}, \ and\ \bibinfo {author} {\bibfnamefont {P.}~\bibnamefont {Simon}},\
  }\href {\doibase 10.1038/nenergy.2016.70} {\bibfield  {journal} {\bibinfo
  {journal} {Nature Energy}\ }\textbf {\bibinfo {volume} {1}},\ \bibinfo
  {pages} {16070} (\bibinfo {year} {2016})}\BibitemShut {NoStop}%
\bibitem [{\citenamefont {Siria}\ \emph {et~al.}(2017)\citenamefont {Siria},
  \citenamefont {Bocquet},\ and\ \citenamefont {Bocquet}}]{siria_new_2017}%
  \BibitemOpen
  \bibfield  {author} {\bibinfo {author} {\bibfnamefont {A.}~\bibnamefont
  {Siria}}, \bibinfo {author} {\bibfnamefont {M.-L.}\ \bibnamefont {Bocquet}},
  \ and\ \bibinfo {author} {\bibfnamefont {L.}~\bibnamefont {Bocquet}},\ }\href
  {\doibase 10.1038/s41570-017-0091} {\bibfield  {journal} {\bibinfo  {journal}
  {Nature Reviews Chemistry}\ }\textbf {\bibinfo {volume} {1}},\ \bibinfo
  {pages} {0091} (\bibinfo {year} {2017})}\BibitemShut {NoStop}%
\bibitem [{\citenamefont {McCaffrey}\ \emph {et~al.}(2017)\citenamefont
  {McCaffrey}, \citenamefont {Nguyen}, \citenamefont {Cox}, \citenamefont
  {Weller}, \citenamefont {Alivisatos}, \citenamefont {Geissler},\ and\
  \citenamefont {Saykally}}]{mccaffrey_mechanism_2017}%
  \BibitemOpen
  \bibfield  {author} {\bibinfo {author} {\bibfnamefont {D.~L.}\ \bibnamefont
  {McCaffrey}}, \bibinfo {author} {\bibfnamefont {S.~C.}\ \bibnamefont
  {Nguyen}}, \bibinfo {author} {\bibfnamefont {S.~J.}\ \bibnamefont {Cox}},
  \bibinfo {author} {\bibfnamefont {H.}~\bibnamefont {Weller}}, \bibinfo
  {author} {\bibfnamefont {A.~P.}\ \bibnamefont {Alivisatos}}, \bibinfo
  {author} {\bibfnamefont {P.~L.}\ \bibnamefont {Geissler}}, \ and\ \bibinfo
  {author} {\bibfnamefont {R.~J.}\ \bibnamefont {Saykally}},\ }\href {\doibase
  10.1073/pnas.1702760114} {\bibfield  {journal} {\bibinfo  {journal}
  {Proceedings of the National Academy of Sciences}\ }\textbf {\bibinfo
  {volume} {114}},\ \bibinfo {pages} {13369} (\bibinfo {year}
  {2017})}\BibitemShut {NoStop}%
\bibitem [{\citenamefont {Iamprasertkun}\ \emph {et~al.}(2019)\citenamefont
  {Iamprasertkun}, \citenamefont {Hirunpinyopas}, \citenamefont {Keerthi},
  \citenamefont {Wang}, \citenamefont {Radha}, \citenamefont {Bissett},\ and\
  \citenamefont {Dryfe}}]{iamprasertkun_capacitance_2019}%
  \BibitemOpen
  \bibfield  {author} {\bibinfo {author} {\bibfnamefont {P.}~\bibnamefont
  {Iamprasertkun}}, \bibinfo {author} {\bibfnamefont {W.}~\bibnamefont
  {Hirunpinyopas}}, \bibinfo {author} {\bibfnamefont {A.}~\bibnamefont
  {Keerthi}}, \bibinfo {author} {\bibfnamefont {B.}~\bibnamefont {Wang}},
  \bibinfo {author} {\bibfnamefont {B.}~\bibnamefont {Radha}}, \bibinfo
  {author} {\bibfnamefont {M.~A.}\ \bibnamefont {Bissett}}, \ and\ \bibinfo
  {author} {\bibfnamefont {R.~A.~W.}\ \bibnamefont {Dryfe}},\ }\href {\doibase
  10.1021/acs.jpclett.8b03523} {\bibfield  {journal} {\bibinfo  {journal} {The
  Journal of Physical Chemistry Letters}\ }\textbf {\bibinfo {volume} {10}},\
  \bibinfo {pages} {617} (\bibinfo {year} {2019})}\BibitemShut {NoStop}%
\bibitem [{\citenamefont {Misra}\ and\ \citenamefont
  {Blankschtein}(2021{\natexlab{a}})}]{misra_ion_2021}%
  \BibitemOpen
  \bibfield  {author} {\bibinfo {author} {\bibfnamefont {R.~P.}\ \bibnamefont
  {Misra}}\ and\ \bibinfo {author} {\bibfnamefont {D.}~\bibnamefont
  {Blankschtein}},\ }\href {\doibase 10.1021/acs.jpcc.0c09855} {\bibfield
  {journal} {\bibinfo  {journal} {The Journal of Physical Chemistry C}\
  }\textbf {\bibinfo {volume} {125}},\ \bibinfo {pages} {2666} (\bibinfo {year}
  {2021}{\natexlab{a}})}\BibitemShut {NoStop}%
\bibitem [{\citenamefont {Scalfi}\ \emph {et~al.}(2021)\citenamefont {Scalfi},
  \citenamefont {Salanne},\ and\ \citenamefont
  {Rotenberg}}]{scalfi_molecular_2021}%
  \BibitemOpen
  \bibfield  {author} {\bibinfo {author} {\bibfnamefont {L.}~\bibnamefont
  {Scalfi}}, \bibinfo {author} {\bibfnamefont {M.}~\bibnamefont {Salanne}}, \
  and\ \bibinfo {author} {\bibfnamefont {B.}~\bibnamefont {Rotenberg}},\ }\href
  {https://doi.org/10.1146/annurev-physchem-090519-024042} {\bibfield
  {journal} {\bibinfo  {journal} {Annual Review of Physical Chemistry}\
  }\textbf {\bibinfo {volume} {72}},\ \bibinfo {pages} {189} (\bibinfo {year}
  {2021})}\BibitemShut {NoStop}%
\bibitem [{\citenamefont {Williams}\ \emph {et~al.}(2017)\citenamefont
  {Williams}, \citenamefont {Dix}, \citenamefont {Troisi},\ and\ \citenamefont
  {Carbone}}]{williams_effective_2017}%
  \BibitemOpen
  \bibfield  {author} {\bibinfo {author} {\bibfnamefont {C.~D.}\ \bibnamefont
  {Williams}}, \bibinfo {author} {\bibfnamefont {J.}~\bibnamefont {Dix}},
  \bibinfo {author} {\bibfnamefont {A.}~\bibnamefont {Troisi}}, \ and\ \bibinfo
  {author} {\bibfnamefont {P.}~\bibnamefont {Carbone}},\ }\href {\doibase
  10.1021/acs.jpclett.6b02783} {\bibfield  {journal} {\bibinfo  {journal} {The
  Journal of Physical Chemistry Letters}\ }\textbf {\bibinfo {volume} {8}},\
  \bibinfo {pages} {703} (\bibinfo {year} {2017})}\BibitemShut {NoStop}%
\bibitem [{\citenamefont {Zhan}\ \emph {et~al.}(2019)\citenamefont {Zhan},
  \citenamefont {Ceron}, \citenamefont {Hawks}, \citenamefont {Otani},
  \citenamefont {Wood}, \citenamefont {Pham}, \citenamefont {Stadermann},\ and\
  \citenamefont {Campbell}}]{zhan_specific_2019}%
  \BibitemOpen
  \bibfield  {author} {\bibinfo {author} {\bibfnamefont {C.}~\bibnamefont
  {Zhan}}, \bibinfo {author} {\bibfnamefont {M.~R.}\ \bibnamefont {Ceron}},
  \bibinfo {author} {\bibfnamefont {S.~A.}\ \bibnamefont {Hawks}}, \bibinfo
  {author} {\bibfnamefont {M.}~\bibnamefont {Otani}}, \bibinfo {author}
  {\bibfnamefont {B.~C.}\ \bibnamefont {Wood}}, \bibinfo {author}
  {\bibfnamefont {T.~A.}\ \bibnamefont {Pham}}, \bibinfo {author}
  {\bibfnamefont {M.}~\bibnamefont {Stadermann}}, \ and\ \bibinfo {author}
  {\bibfnamefont {P.~G.}\ \bibnamefont {Campbell}},\ }\href {\doibase
  10.1038/s41467-019-12854-7} {\bibfield  {journal} {\bibinfo  {journal}
  {Nature Communications}\ }\textbf {\bibinfo {volume} {10}},\ \bibinfo {pages}
  {4858} (\bibinfo {year} {2019})}\BibitemShut {NoStop}%
\bibitem [{\citenamefont {Ruggeri}\ \emph {et~al.}(2022)\citenamefont
  {Ruggeri}, \citenamefont {Reeves}, \citenamefont {Hsu}, \citenamefont
  {Jeanmairet}, \citenamefont {Salanne},\ and\ \citenamefont
  {Pierleoni}}]{ruggeri_multi-scale_2021}%
  \BibitemOpen
  \bibfield  {author} {\bibinfo {author} {\bibfnamefont {M.}~\bibnamefont
  {Ruggeri}}, \bibinfo {author} {\bibfnamefont {K.}~\bibnamefont {Reeves}},
  \bibinfo {author} {\bibfnamefont {T.-Y.}\ \bibnamefont {Hsu}}, \bibinfo
  {author} {\bibfnamefont {G.}~\bibnamefont {Jeanmairet}}, \bibinfo {author}
  {\bibfnamefont {M.}~\bibnamefont {Salanne}}, \ and\ \bibinfo {author}
  {\bibfnamefont {C.}~\bibnamefont {Pierleoni}},\ }\href
  {https://doi.org/10.1063/5.0082944} {\bibfield  {journal} {\bibinfo
  {journal} {The Journal of Chemical Physics}\ }\textbf {\bibinfo {volume}
  {156}},\ \bibinfo {pages} {094709} (\bibinfo {year} {2022})}\BibitemShut
  {NoStop}%
\bibitem [{\citenamefont {Grosjean}\ \emph {et~al.}(2019)\citenamefont
  {Grosjean}, \citenamefont {Bocquet},\ and\ \citenamefont
  {Vuilleumier}}]{grosjean_versatile_2019}%
  \BibitemOpen
  \bibfield  {author} {\bibinfo {author} {\bibfnamefont {B.}~\bibnamefont
  {Grosjean}}, \bibinfo {author} {\bibfnamefont {M.-L.}\ \bibnamefont
  {Bocquet}}, \ and\ \bibinfo {author} {\bibfnamefont {R.}~\bibnamefont
  {Vuilleumier}},\ }\href {\doibase 10.1038/s41467-019-09708-7} {\bibfield
  {journal} {\bibinfo  {journal} {Nature Communications}\ }\textbf {\bibinfo
  {volume} {10}},\ \bibinfo {pages} {1656} (\bibinfo {year}
  {2019})}\BibitemShut {NoStop}%
\bibitem [{\citenamefont {Joly}\ \emph {et~al.}(2021)\citenamefont {Joly},
  \citenamefont {Mei{\ss}ner}, \citenamefont {Iannuzzi},\ and\ \citenamefont
  {Tocci}}]{joly_osmotic_2021}%
  \BibitemOpen
  \bibfield  {author} {\bibinfo {author} {\bibfnamefont {L.}~\bibnamefont
  {Joly}}, \bibinfo {author} {\bibfnamefont {R.~H.}\ \bibnamefont
  {Mei{\ss}ner}}, \bibinfo {author} {\bibfnamefont {M.}~\bibnamefont
  {Iannuzzi}}, \ and\ \bibinfo {author} {\bibfnamefont {G.}~\bibnamefont
  {Tocci}},\ }\href {https://doi.org/10.1021/acsnano.1c05931} {\bibfield
  {journal} {\bibinfo  {journal} {ACS Nano}\ }\textbf {\bibinfo {volume}
  {15}},\ \bibinfo {pages} {15249} (\bibinfo {year} {2021})}\BibitemShut
  {NoStop}%
\bibitem [{\citenamefont {Son}\ and\ \citenamefont
  {Wang}(2021)}]{son_image-charge_2021}%
  \BibitemOpen
  \bibfield  {author} {\bibinfo {author} {\bibfnamefont {C.~Y.}\ \bibnamefont
  {Son}}\ and\ \bibinfo {author} {\bibfnamefont {Z.-G.}\ \bibnamefont {Wang}},\
  }\href {\doibase 10.1073/pnas.2020615118} {\bibfield  {journal} {\bibinfo
  {journal} {Proceedings of the National Academy of Sciences}\ }\textbf
  {\bibinfo {volume} {118}},\ \bibinfo {pages} {e2020615118} (\bibinfo {year}
  {2021})}\BibitemShut {NoStop}%
\bibitem [{\citenamefont {Scalfi}\ \emph {et~al.}(2020)\citenamefont {Scalfi},
  \citenamefont {Dufils}, \citenamefont {Reeves}, \citenamefont {Rotenberg},\
  and\ \citenamefont {Salanne}}]{scalfi_semiclassical_2020}%
  \BibitemOpen
  \bibfield  {author} {\bibinfo {author} {\bibfnamefont {L.}~\bibnamefont
  {Scalfi}}, \bibinfo {author} {\bibfnamefont {T.}~\bibnamefont {Dufils}},
  \bibinfo {author} {\bibfnamefont {K.~G.}\ \bibnamefont {Reeves}}, \bibinfo
  {author} {\bibfnamefont {B.}~\bibnamefont {Rotenberg}}, \ and\ \bibinfo
  {author} {\bibfnamefont {M.}~\bibnamefont {Salanne}},\ }\href
  {http://aip.scitation.org/doi/10.1063/5.0028232} {\bibfield  {journal}
  {\bibinfo  {journal} {The Journal of Chemical Physics}\ }\textbf {\bibinfo
  {volume} {153}},\ \bibinfo {pages} {174704} (\bibinfo {year}
  {2020})}\BibitemShut {NoStop}%
\bibitem [{\citenamefont {Schlaich}\ \emph {et~al.}(2022)\citenamefont
  {Schlaich}, \citenamefont {Jin}, \citenamefont {Bocquet},\ and\ \citenamefont
  {Coasne}}]{schlaich_electronic_2022}%
  \BibitemOpen
  \bibfield  {author} {\bibinfo {author} {\bibfnamefont {A.}~\bibnamefont
  {Schlaich}}, \bibinfo {author} {\bibfnamefont {D.}~\bibnamefont {Jin}},
  \bibinfo {author} {\bibfnamefont {L.}~\bibnamefont {Bocquet}}, \ and\
  \bibinfo {author} {\bibfnamefont {B.}~\bibnamefont {Coasne}},\ }\href
  {https://www.nature.com/articles/s41563-021-01121-0} {\bibfield  {journal}
  {\bibinfo  {journal} {Nature Materials}\ }\textbf {\bibinfo {volume} {21}},\
  \bibinfo {pages} {237} (\bibinfo {year} {2022})}\BibitemShut {NoStop}%
\bibitem [{\citenamefont {Misra}\ and\ \citenamefont
  {Blankschtein}(2021{\natexlab{b}})}]{misra_uncovering_2021}%
  \BibitemOpen
  \bibfield  {author} {\bibinfo {author} {\bibfnamefont {R.~P.}\ \bibnamefont
  {Misra}}\ and\ \bibinfo {author} {\bibfnamefont {D.}~\bibnamefont
  {Blankschtein}},\ }\href {\doibase 10.1021/acs.langmuir.0c02829} {\bibfield
  {journal} {\bibinfo  {journal} {Langmuir}\ }\textbf {\bibinfo {volume}
  {37}},\ \bibinfo {pages} {722} (\bibinfo {year}
  {2021}{\natexlab{b}})}\BibitemShut {NoStop}%
\bibitem [{\citenamefont {Schwinger}(1998)}]{schwinger_chapter_1998}%
  \BibitemOpen
  \bibfield  {author} {\bibinfo {author} {\bibfnamefont {J.}~\bibnamefont
  {Schwinger}},\ }in\ \href@noop {} {\emph {\bibinfo {booktitle} {Classical
  {Electrodynamics}}}}\ (\bibinfo  {publisher} {Westview Press},\ \bibinfo
  {year} {1998})\BibitemShut {NoStop}%
\bibitem [{\citenamefont {Loche}\ \emph {et~al.}(2018)\citenamefont {Loche},
  \citenamefont {Ayaz}, \citenamefont {Schlaich}, \citenamefont {Bonthuis},\
  and\ \citenamefont {Netz}}]{loche_breakdown_2018}%
  \BibitemOpen
  \bibfield  {author} {\bibinfo {author} {\bibfnamefont {P.}~\bibnamefont
  {Loche}}, \bibinfo {author} {\bibfnamefont {C.}~\bibnamefont {Ayaz}},
  \bibinfo {author} {\bibfnamefont {A.}~\bibnamefont {Schlaich}}, \bibinfo
  {author} {\bibfnamefont {D.~J.}\ \bibnamefont {Bonthuis}}, \ and\ \bibinfo
  {author} {\bibfnamefont {R.~R.}\ \bibnamefont {Netz}},\ }\href
  {https://doi.org/10.1021/acs.jpclett.8b02473} {\bibfield  {journal} {\bibinfo
   {journal} {The Journal of Physical Chemistry Letters}\ }\textbf {\bibinfo
  {volume} {9}},\ \bibinfo {pages} {6463} (\bibinfo {year} {2018})}\BibitemShut
  {NoStop}%
\bibitem [{\citenamefont {Vorotyntsev}\ and\ \citenamefont
  {Kornyshev}(1980)}]{vorotyntsev_electrostatic_1980}%
  \BibitemOpen
  \bibfield  {author} {\bibinfo {author} {\bibfnamefont {M.~A.}\ \bibnamefont
  {Vorotyntsev}}\ and\ \bibinfo {author} {\bibfnamefont {A.~A.}\ \bibnamefont
  {Kornyshev}},\ }\href@noop {} {\bibfield  {journal} {\bibinfo  {journal} {Zh.
  Eksp. Teor. Fiz.}\ }\textbf {\bibinfo {volume} {72}},\ \bibinfo {pages}
  {1008} (\bibinfo {year} {1980})}\BibitemShut {NoStop}%
\bibitem [{\citenamefont {Kornyshev}\ and\ \citenamefont
  {Vorotyntsev}(1980)}]{kornyshev_nonlocal_1980}%
  \BibitemOpen
  \bibfield  {author} {\bibinfo {author} {\bibfnamefont {A.~A.}\ \bibnamefont
  {Kornyshev}}\ and\ \bibinfo {author} {\bibfnamefont {M.~A.}\ \bibnamefont
  {Vorotyntsev}},\ }\href
  {https://www.sciencedirect.com/science/article/abs/pii/003960288090597X}
  {\bibfield  {journal} {\bibinfo  {journal} {Surface Science}\ }\textbf
  {\bibinfo {volume} {101}},\ \bibinfo {pages} {23} (\bibinfo {year}
  {1980})}\BibitemShut {NoStop}%
\bibitem [{\citenamefont {Gabovich}\ \emph {et~al.}(2012)\citenamefont
  {Gabovich}, \citenamefont {Li}, \citenamefont {Szymczak},\ and\ \citenamefont
  {Voitenko}}]{gabovich_image_2012}%
  \BibitemOpen
  \bibfield  {author} {\bibinfo {author} {\bibfnamefont {A.~M.}\ \bibnamefont
  {Gabovich}}, \bibinfo {author} {\bibfnamefont {M.~S.}\ \bibnamefont {Li}},
  \bibinfo {author} {\bibfnamefont {H.}~\bibnamefont {Szymczak}}, \ and\
  \bibinfo {author} {\bibfnamefont {A.~I.}\ \bibnamefont {Voitenko}},\ }\href
  {\doibase 10.1016/j.susc.2011.11.020} {\bibfield  {journal} {\bibinfo
  {journal} {Surface Science}\ }\textbf {\bibinfo {volume} {606}},\ \bibinfo
  {pages} {510} (\bibinfo {year} {2012})}\BibitemShut {NoStop}%
\bibitem [{\citenamefont {Li}\ \emph {et~al.}(2012)\citenamefont {Li},
  \citenamefont {Feng}, \citenamefont {Wang}, \citenamefont {Meng},
  \citenamefont {Klime\u{s}},\ and\ \citenamefont
  {Michaelides}}]{li_influence_2012}%
  \BibitemOpen
  \bibfield  {author} {\bibinfo {author} {\bibfnamefont {X.}~\bibnamefont
  {Li}}, \bibinfo {author} {\bibfnamefont {J.}~\bibnamefont {Feng}}, \bibinfo
  {author} {\bibfnamefont {E.}~\bibnamefont {Wang}}, \bibinfo {author}
  {\bibfnamefont {S.}~\bibnamefont {Meng}}, \bibinfo {author} {\bibfnamefont
  {J.}~\bibnamefont {Klime\u{s}}}, \ and\ \bibinfo {author} {\bibfnamefont
  {A.}~\bibnamefont {Michaelides}},\ }\href {\doibase
  10.1103/PhysRevB.85.085425} {\bibfield  {journal} {\bibinfo  {journal}
  {Physical Review B}\ }\textbf {\bibinfo {volume} {85}},\ \bibinfo {pages}
  {085425} (\bibinfo {year} {2012})}\BibitemShut {NoStop}%
\bibitem [{\citenamefont {Berendsen}\ \emph {et~al.}(1987)\citenamefont
  {Berendsen}, \citenamefont {Grigera},\ and\ \citenamefont
  {Straatsma}}]{berendsen_missing_1987}%
  \BibitemOpen
  \bibfield  {author} {\bibinfo {author} {\bibfnamefont {H.~J.~C.}\
  \bibnamefont {Berendsen}}, \bibinfo {author} {\bibfnamefont {J.~R.}\
  \bibnamefont {Grigera}}, \ and\ \bibinfo {author} {\bibfnamefont {T.~P.}\
  \bibnamefont {Straatsma}},\ }\href {https://doi.org/10.1021/j100308a038}
  {\bibfield  {journal} {\bibinfo  {journal} {The Journal of Physical
  Chemistry}\ }\textbf {\bibinfo {volume} {91}},\ \bibinfo {pages} {6269}
  (\bibinfo {year} {1987})}\BibitemShut {NoStop}%
\bibitem [{\citenamefont {Jeanmairet}\ \emph {et~al.}(2016)\citenamefont
  {Jeanmairet}, \citenamefont {Levy}, \citenamefont {Levesque},\ and\
  \citenamefont {Borgis}}]{jeanmairet_molecular_2016}%
  \BibitemOpen
  \bibfield  {author} {\bibinfo {author} {\bibfnamefont {G.}~\bibnamefont
  {Jeanmairet}}, \bibinfo {author} {\bibfnamefont {N.}~\bibnamefont {Levy}},
  \bibinfo {author} {\bibfnamefont {M.}~\bibnamefont {Levesque}}, \ and\
  \bibinfo {author} {\bibfnamefont {D.}~\bibnamefont {Borgis}},\ }\href
  {\doibase 10.1088/0953-8984/28/24/244005} {\bibfield  {journal} {\bibinfo
  {journal} {Journal of Physics: Condensed Matter}\ }\textbf {\bibinfo {volume}
  {28}},\ \bibinfo {pages} {244005} (\bibinfo {year} {2016})}\BibitemShut
  {NoStop}%
\bibitem [{\citenamefont {Bopp}\ \emph {et~al.}(1996)\citenamefont {Bopp},
  \citenamefont {Kornyshev},\ and\ \citenamefont {Sutmann}}]{bopp_static_1996}%
  \BibitemOpen
  \bibfield  {author} {\bibinfo {author} {\bibfnamefont {P.~A.}\ \bibnamefont
  {Bopp}}, \bibinfo {author} {\bibfnamefont {A.~A.}\ \bibnamefont {Kornyshev}},
  \ and\ \bibinfo {author} {\bibfnamefont {G.}~\bibnamefont {Sutmann}},\ }\href
  {\doibase 10.1103/PhysRevLett.76.1280} {\bibfield  {journal} {\bibinfo
  {journal} {Physical Review Letters}\ }\textbf {\bibinfo {volume} {76}},\
  \bibinfo {pages} {1280} (\bibinfo {year} {1996})}\BibitemShut {NoStop}%
\bibitem [{\citenamefont {Hansen}\ and\ \citenamefont
  {McDonald}(2013)}]{hansen_theory_2013}%
  \BibitemOpen
  \bibfield  {author} {\bibinfo {author} {\bibfnamefont {J.-P.}\ \bibnamefont
  {Hansen}}\ and\ \bibinfo {author} {\bibfnamefont {I.~R.}\ \bibnamefont
  {McDonald}},\ }\href@noop {} {\emph {\bibinfo {title} {Theory of simple
  liquids}}}\ (\bibinfo  {publisher} {Academic Press},\ \bibinfo {year}
  {2013})\BibitemShut {NoStop}%
\bibitem [{\citenamefont {Bonthuis}\ \emph {et~al.}(2012)\citenamefont
  {Bonthuis}, \citenamefont {Gekle},\ and\ \citenamefont
  {Netz}}]{bonthuis_profile_2012}%
  \BibitemOpen
  \bibfield  {author} {\bibinfo {author} {\bibfnamefont {D.~J.}\ \bibnamefont
  {Bonthuis}}, \bibinfo {author} {\bibfnamefont {S.}~\bibnamefont {Gekle}}, \
  and\ \bibinfo {author} {\bibfnamefont {R.~R.}\ \bibnamefont {Netz}},\ }\href
  {\doibase 10.1021/la2051564} {\bibfield  {journal} {\bibinfo  {journal}
  {Langmuir}\ }\textbf {\bibinfo {volume} {28}},\ \bibinfo {pages} {7679}
  (\bibinfo {year} {2012})}\BibitemShut {NoStop}%
\bibitem [{\citenamefont {Castro~Neto}\ \emph {et~al.}(2009)\citenamefont
  {Castro~Neto}, \citenamefont {Guinea}, \citenamefont {Peres}, \citenamefont
  {Novoselov},\ and\ \citenamefont {Geim}}]{castro_neto_electronic_2009}%
  \BibitemOpen
  \bibfield  {author} {\bibinfo {author} {\bibfnamefont {A.~H.}\ \bibnamefont
  {Castro~Neto}}, \bibinfo {author} {\bibfnamefont {F.}~\bibnamefont {Guinea}},
  \bibinfo {author} {\bibfnamefont {N.~M.~R.}\ \bibnamefont {Peres}}, \bibinfo
  {author} {\bibfnamefont {K.~S.}\ \bibnamefont {Novoselov}}, \ and\ \bibinfo
  {author} {\bibfnamefont {A.~K.}\ \bibnamefont {Geim}},\ }\href {\doibase
  10.1103/RevModPhys.81.109} {\bibfield  {journal} {\bibinfo  {journal}
  {Reviews of Modern Physics}\ }\textbf {\bibinfo {volume} {81}},\ \bibinfo
  {pages} {109} (\bibinfo {year} {2009})}\BibitemShut {NoStop}%
\bibitem [{\citenamefont {Mahan}(1990)}]{mahan_many-particle_1990}%
  \BibitemOpen
  \bibfield  {author} {\bibinfo {author} {\bibfnamefont {G.~D.}\ \bibnamefont
  {Mahan}},\ }\href@noop {} {\emph {\bibinfo {title} {Many-{Particle}
  {Physics}}}}\ (\bibinfo  {publisher} {Springer US},\ \bibinfo {year}
  {1990})\BibitemShut {NoStop}%
\bibitem [{\citenamefont {Hwang}\ and\ \citenamefont
  {Das~Sarma}(2007)}]{hwang_dielectric_2007}%
  \BibitemOpen
  \bibfield  {author} {\bibinfo {author} {\bibfnamefont {E.~H.}\ \bibnamefont
  {Hwang}}\ and\ \bibinfo {author} {\bibfnamefont {S.}~\bibnamefont
  {Das~Sarma}},\ }\href {\doibase 10.1103/PhysRevB.75.205418} {\bibfield
  {journal} {\bibinfo  {journal} {Physical Review B}\ }\textbf {\bibinfo
  {volume} {75}},\ \bibinfo {pages} {205418} (\bibinfo {year}
  {2007})}\BibitemShut {NoStop}%
\bibitem [{\citenamefont {Bohm}\ and\ \citenamefont
  {Pines}(1953)}]{bohm_collective_1953}%
  \BibitemOpen
  \bibfield  {author} {\bibinfo {author} {\bibfnamefont {D.}~\bibnamefont
  {Bohm}}\ and\ \bibinfo {author} {\bibfnamefont {D.}~\bibnamefont {Pines}},\
  }\href {https://link.aps.org/doi/10.1103/PhysRev.92.609} {\bibfield
  {journal} {\bibinfo  {journal} {Physical Review}\ }\textbf {\bibinfo {volume}
  {92}},\ \bibinfo {pages} {609} (\bibinfo {year} {1953})}\BibitemShut
  {NoStop}%
\bibitem [{\citenamefont {Kaiser}\ \emph {et~al.}(2017)\citenamefont {Kaiser},
  \citenamefont {Comtet}, \citenamefont {Niguès}, \citenamefont {Siria},
  \citenamefont {Coasne},\ and\ \citenamefont
  {Bocquet}}]{kaiser_electrostatic_2017}%
  \BibitemOpen
  \bibfield  {author} {\bibinfo {author} {\bibfnamefont {V.}~\bibnamefont
  {Kaiser}}, \bibinfo {author} {\bibfnamefont {J.}~\bibnamefont {Comtet}},
  \bibinfo {author} {\bibfnamefont {A.}~\bibnamefont {Niguès}}, \bibinfo
  {author} {\bibfnamefont {A.}~\bibnamefont {Siria}}, \bibinfo {author}
  {\bibfnamefont {B.}~\bibnamefont {Coasne}}, \ and\ \bibinfo {author}
  {\bibfnamefont {L.}~\bibnamefont {Bocquet}},\ }\href {\doibase
  10.1039/C6FD00256K} {\bibfield  {journal} {\bibinfo  {journal} {Faraday
  Discussions}\ }\textbf {\bibinfo {volume} {199}},\ \bibinfo {pages} {129}
  (\bibinfo {year} {2017})}\BibitemShut {NoStop}%
\bibitem [{Note1()}]{Note1}%
  \BibitemOpen
  \bibinfo {note} {Using $N_{\protect \text {ext}}[z]=J_{0}(q\protect \sqrt
  {b^{2}+(z-z_{0})^{2}})/2b$ with $J_{0}$ being the zeroth order Bessel
  function, we compute $F(z_{0})=\protect \frac {1}{2}\DOTSI \intop \ilimits@
  _{0}^{+\infty }\protect \frac {\protect \text {d}q}{2\pi }q\left [N_{\protect
  \text {ext}}^{\dagger }(W-V)N_{\protect \text {ext}}\right ](q),$ for the
  three Green's functions, in log-log space using $q=e^{y}E_{F}/v_{F}$, $y\in
  [-1,8]$ and $N_{y}=100$ for convergence. Note that $V[z,z']=e^{-q\delimiter
  "026A30C z-z'\delimiter "026A30C }/2\epsilon _{0}q$.}\BibitemShut {Stop}%
\bibitem [{\citenamefont {Scalfi}\ and\ \citenamefont
  {Rotenberg}(2021)}]{scalfi_microscopic_2021}%
  \BibitemOpen
  \bibfield  {author} {\bibinfo {author} {\bibfnamefont {L.}~\bibnamefont
  {Scalfi}}\ and\ \bibinfo {author} {\bibfnamefont {B.}~\bibnamefont
  {Rotenberg}},\ }\href {https://www.pnas.org/content/118/50/e2108769118}
  {\bibfield  {journal} {\bibinfo  {journal} {Proceedings of the National
  Academy of Sciences}\ }\textbf {\bibinfo {volume} {118}},\ \bibinfo {pages}
  {e2108769118} (\bibinfo {year} {2021})}\BibitemShut {NoStop}%
\bibitem [{\citenamefont {Marcus}(2009)}]{marcus_effect_2009}%
  \BibitemOpen
  \bibfield  {author} {\bibinfo {author} {\bibfnamefont {Y.}~\bibnamefont
  {Marcus}},\ }\href {\doibase 10.1021/cr8003828} {\bibfield  {journal}
  {\bibinfo  {journal} {Chemical Reviews}\ }\textbf {\bibinfo {volume} {109}},\
  \bibinfo {pages} {1346} (\bibinfo {year} {2009})}\BibitemShut {NoStop}%
\end{thebibliography}%


\begin{thebibliography}{17}%
\makeatletter
\providecommand \@ifxundefined [1]{%
 \@ifx{#1\undefined}
}%
\providecommand \@ifnum [1]{%
 \ifnum #1\expandafter \@firstoftwo
 \else \expandafter \@secondoftwo
 \fi
}%
\providecommand \@ifx [1]{%
 \ifx #1\expandafter \@firstoftwo
 \else \expandafter \@secondoftwo
 \fi
}%
\providecommand \natexlab [1]{#1}%
\providecommand \enquote  [1]{``#1''}%
\providecommand \bibnamefont  [1]{#1}%
\providecommand \bibfnamefont [1]{#1}%
\providecommand \citenamefont [1]{#1}%
\providecommand \href@noop [0]{\@secondoftwo}%
\providecommand \href [0]{\begingroup \@sanitize@url \@href}%
\providecommand \@href[1]{\@@startlink{#1}\@@href}%
\providecommand \@@href[1]{\endgroup#1\@@endlink}%
\providecommand \@sanitize@url [0]{\catcode `\\12\catcode `\$12\catcode
  `\&12\catcode `\#12\catcode `\^12\catcode `\_12\catcode `\%12\relax}%
\providecommand \@@startlink[1]{}%
\providecommand \@@endlink[0]{}%
\providecommand \url  [0]{\begingroup\@sanitize@url \@url }%
\providecommand \@url [1]{\endgroup\@href {#1}{\urlprefix }}%
\providecommand \urlprefix  [0]{URL }%
\providecommand \Eprint [0]{\href }%
\providecommand \doibase [0]{http://dx.doi.org/}%
\providecommand \selectlanguage [0]{\@gobble}%
\providecommand \bibinfo  [0]{\@secondoftwo}%
\providecommand \bibfield  [0]{\@secondoftwo}%
\providecommand \translation [1]{[#1]}%
\providecommand \BibitemOpen [0]{}%
\providecommand \bibitemStop [0]{}%
\providecommand \bibitemNoStop [0]{.\EOS\space}%
\providecommand \EOS [0]{\spacefactor3000\relax}%
\providecommand \BibitemShut  [1]{\csname bibitem#1\endcsname}%
\let\auto@bib@innerbib\@empty
%</preamble>
\bibitem [{\citenamefont {Schwinger}(1998)}]{schwinger_chapter_1998}%
  \BibitemOpen
  \bibfield  {author} {\bibinfo {author} {\bibfnamefont {J.}~\bibnamefont
  {Schwinger}},\ }in\ \href@noop {} {\emph {\bibinfo {booktitle} {Classical
  {Electrodynamics}}}}\ (\bibinfo  {publisher} {Westview Press},\ \bibinfo
  {year} {1998})\BibitemShut {NoStop}%
\bibitem [{\citenamefont {Mahan}(1990)}]{mahan_many-particle_1990}%
  \BibitemOpen
  \bibfield  {author} {\bibinfo {author} {\bibfnamefont {G.~D.}\ \bibnamefont
  {Mahan}},\ }\href@noop {} {\emph {\bibinfo {title} {Many-{Particle}
  {Physics}}}}\ (\bibinfo  {publisher} {Springer US},\ \bibinfo {year}
  {1990})\BibitemShut {NoStop}%
\bibitem [{\citenamefont {Hansen}\ and\ \citenamefont
  {McDonald}(2013)}]{hansen_theory_2013}%
  \BibitemOpen
  \bibfield  {author} {\bibinfo {author} {\bibfnamefont {J.-P.}\ \bibnamefont
  {Hansen}}\ and\ \bibinfo {author} {\bibfnamefont {I.~R.}\ \bibnamefont
  {McDonald}},\ }\href@noop {} {\emph {\bibinfo {title} {Theory of simple
  liquids}}}\ (\bibinfo  {publisher} {Academic Press},\ \bibinfo {year}
  {2013})\BibitemShut {NoStop}%
\bibitem [{\citenamefont {Bohm}\ and\ \citenamefont
  {Pines}(1953)}]{bohm_collective_1953}%
  \BibitemOpen
  \bibfield  {author} {\bibinfo {author} {\bibfnamefont {D.}~\bibnamefont
  {Bohm}}\ and\ \bibinfo {author} {\bibfnamefont {D.}~\bibnamefont {Pines}},\
  }\href {https://link.aps.org/doi/10.1103/PhysRev.92.609} {\bibfield
  {journal} {\bibinfo  {journal} {Physical Review}\ }\textbf {\bibinfo {volume}
  {92}},\ \bibinfo {pages} {609} (\bibinfo {year} {1953})}\BibitemShut
  {NoStop}%
\bibitem [{\citenamefont {Kavokine}\ \emph {et~al.}(2022)\citenamefont
  {Kavokine}, \citenamefont {Bocquet},\ and\ \citenamefont
  {Bocquet}}]{kavokine_fluctuation-induced_2022}%
  \BibitemOpen
  \bibfield  {author} {\bibinfo {author} {\bibfnamefont {N.}~\bibnamefont
  {Kavokine}}, \bibinfo {author} {\bibfnamefont {M.-L.}\ \bibnamefont
  {Bocquet}}, \ and\ \bibinfo {author} {\bibfnamefont {L.}~\bibnamefont
  {Bocquet}},\ }\href {https://www.nature.com/articles/s41586-021-04284-7}
  {\bibfield  {journal} {\bibinfo  {journal} {Nature}\ }\textbf {\bibinfo
  {volume} {602}},\ \bibinfo {pages} {84} (\bibinfo {year} {2022})}\BibitemShut
  {NoStop}%
\bibitem [{\citenamefont {Hedin}\ and\ \citenamefont
  {Lundqvist}(1970)}]{hedin_effects_1970}%
  \BibitemOpen
  \bibfield  {author} {\bibinfo {author} {\bibfnamefont {L.}~\bibnamefont
  {Hedin}}\ and\ \bibinfo {author} {\bibfnamefont {S.}~\bibnamefont
  {Lundqvist}},\ }\href
  {https://www.sciencedirect.com/science/article/pii/S0081194708606153}
  {\bibfield  {journal} {\bibinfo  {journal} {Solid State Physics}\ }\textbf
  {\bibinfo {volume} {23}},\ \bibinfo {pages} {1} (\bibinfo {year}
  {1970})}\BibitemShut {NoStop}%
\bibitem [{\citenamefont {Giustino}(2017)}]{giustino_electron-phonon_2017}%
  \BibitemOpen
  \bibfield  {author} {\bibinfo {author} {\bibfnamefont {F.}~\bibnamefont
  {Giustino}},\ }\href {\doibase 10.1103/RevModPhys.89.015003} {\bibfield
  {journal} {\bibinfo  {journal} {Reviews of Modern Physics}\ }\textbf
  {\bibinfo {volume} {89}},\ \bibinfo {pages} {015003} (\bibinfo {year}
  {2017})}\BibitemShut {NoStop}%
\bibitem [{\citenamefont {Kubo}(1966)}]{kubo_fluctuation-dissipation_1966}%
  \BibitemOpen
  \bibfield  {author} {\bibinfo {author} {\bibfnamefont {R.}~\bibnamefont
  {Kubo}},\ }\href {https://doi.org/10.1088/0034-4885/29/1/306} {\bibfield
  {journal} {\bibinfo  {journal} {Rep. Prog. Phys.}\ }\textbf {\bibinfo
  {volume} {29}} (\bibinfo {year} {1966})}\BibitemShut {NoStop}%
\bibitem [{\citenamefont {Jeanmairet}\ \emph {et~al.}(2013)\citenamefont
  {Jeanmairet}, \citenamefont {Levesque}, \citenamefont {Vuilleumier},\ and\
  \citenamefont {Borgis}}]{jeanmairet_molecular_2013}%
  \BibitemOpen
  \bibfield  {author} {\bibinfo {author} {\bibfnamefont {G.}~\bibnamefont
  {Jeanmairet}}, \bibinfo {author} {\bibfnamefont {M.}~\bibnamefont
  {Levesque}}, \bibinfo {author} {\bibfnamefont {R.}~\bibnamefont
  {Vuilleumier}}, \ and\ \bibinfo {author} {\bibfnamefont {D.}~\bibnamefont
  {Borgis}},\ }\href@noop {} {\bibfield  {journal} {\bibinfo  {journal} {J.
  Phys. Chem. Lett.}\ ,\ \bibinfo {pages} {6}} (\bibinfo {year}
  {2013})}\BibitemShut {NoStop}%
\bibitem [{\citenamefont {Berendsen}\ \emph {et~al.}(1987)\citenamefont
  {Berendsen}, \citenamefont {Grigera},\ and\ \citenamefont
  {Straatsma}}]{berendsen_missing_1987}%
  \BibitemOpen
  \bibfield  {author} {\bibinfo {author} {\bibfnamefont {H.~J.~C.}\
  \bibnamefont {Berendsen}}, \bibinfo {author} {\bibfnamefont {J.~R.}\
  \bibnamefont {Grigera}}, \ and\ \bibinfo {author} {\bibfnamefont {T.~P.}\
  \bibnamefont {Straatsma}},\ }\href {https://doi.org/10.1021/j100308a038}
  {\bibfield  {journal} {\bibinfo  {journal} {The Journal of Physical
  Chemistry}\ }\textbf {\bibinfo {volume} {91}},\ \bibinfo {pages} {6269}
  (\bibinfo {year} {1987})}\BibitemShut {NoStop}%
\bibitem [{\citenamefont {Jeanmairet}\ \emph {et~al.}(2016)\citenamefont
  {Jeanmairet}, \citenamefont {Levy}, \citenamefont {Levesque},\ and\
  \citenamefont {Borgis}}]{jeanmairet_molecular_2016}%
  \BibitemOpen
  \bibfield  {author} {\bibinfo {author} {\bibfnamefont {G.}~\bibnamefont
  {Jeanmairet}}, \bibinfo {author} {\bibfnamefont {N.}~\bibnamefont {Levy}},
  \bibinfo {author} {\bibfnamefont {M.}~\bibnamefont {Levesque}}, \ and\
  \bibinfo {author} {\bibfnamefont {D.}~\bibnamefont {Borgis}},\ }\href
  {\doibase 10.1088/0953-8984/28/24/244005} {\bibfield  {journal} {\bibinfo
  {journal} {Journal of Physics: Condensed Matter}\ }\textbf {\bibinfo {volume}
  {28}},\ \bibinfo {pages} {244005} (\bibinfo {year} {2016})}\BibitemShut
  {NoStop}%
\bibitem [{\citenamefont {Bopp}\ \emph {et~al.}(1996)\citenamefont {Bopp},
  \citenamefont {Kornyshev},\ and\ \citenamefont {Sutmann}}]{bopp_static_1996}%
  \BibitemOpen
  \bibfield  {author} {\bibinfo {author} {\bibfnamefont {P.~A.}\ \bibnamefont
  {Bopp}}, \bibinfo {author} {\bibfnamefont {A.~A.}\ \bibnamefont {Kornyshev}},
  \ and\ \bibinfo {author} {\bibfnamefont {G.}~\bibnamefont {Sutmann}},\ }\href
  {\doibase 10.1103/PhysRevLett.76.1280} {\bibfield  {journal} {\bibinfo
  {journal} {Physical Review Letters}\ }\textbf {\bibinfo {volume} {76}},\
  \bibinfo {pages} {1280} (\bibinfo {year} {1996})}\BibitemShut {NoStop}%
\bibitem [{\citenamefont {Monet}\ \emph {et~al.}(2021)\citenamefont {Monet},
  \citenamefont {Bresme}, \citenamefont {Kornyshev},\ and\ \citenamefont
  {Berthoumieux}}]{monet_nonlocal_2021}%
  \BibitemOpen
  \bibfield  {author} {\bibinfo {author} {\bibfnamefont {G.}~\bibnamefont
  {Monet}}, \bibinfo {author} {\bibfnamefont {F.}~\bibnamefont {Bresme}},
  \bibinfo {author} {\bibfnamefont {A.}~\bibnamefont {Kornyshev}}, \ and\
  \bibinfo {author} {\bibfnamefont {H.}~\bibnamefont {Berthoumieux}},\ }\href
  {\doibase 10.1103/PhysRevLett.126.216001} {\bibfield  {journal} {\bibinfo
  {journal} {Physical Review Letters}\ }\textbf {\bibinfo {volume} {126}},\
  \bibinfo {pages} {216001} (\bibinfo {year} {2021})}\BibitemShut {NoStop}%
\bibitem [{\citenamefont {Werder}\ \emph {et~al.}(2003)\citenamefont {Werder},
  \citenamefont {Walther}, \citenamefont {Jaffe}, \citenamefont {Halicioglu},\
  and\ \citenamefont {Koumoutsakos}}]{Werder2003}%
  \BibitemOpen
  \bibfield  {author} {\bibinfo {author} {\bibfnamefont {T.}~\bibnamefont
  {Werder}}, \bibinfo {author} {\bibfnamefont {J.~H.}\ \bibnamefont {Walther}},
  \bibinfo {author} {\bibfnamefont {R.~L.}\ \bibnamefont {Jaffe}}, \bibinfo
  {author} {\bibfnamefont {T.}~\bibnamefont {Halicioglu}}, \ and\ \bibinfo
  {author} {\bibfnamefont {P.}~\bibnamefont {Koumoutsakos}},\ }\href {\doibase
  10.1021/jp0268112} {\bibfield  {journal} {\bibinfo  {journal} {The Journal of
  Physical Chemistry B}\ }\textbf {\bibinfo {volume} {107}},\ \bibinfo {pages}
  {1345} (\bibinfo {year} {2003})}\BibitemShut {NoStop}%
\bibitem [{\citenamefont {Hwang}\ and\ \citenamefont
  {Das~Sarma}(2007)}]{hwang_dielectric_2007}%
  \BibitemOpen
  \bibfield  {author} {\bibinfo {author} {\bibfnamefont {E.~H.}\ \bibnamefont
  {Hwang}}\ and\ \bibinfo {author} {\bibfnamefont {S.}~\bibnamefont
  {Das~Sarma}},\ }\href {\doibase 10.1103/PhysRevB.75.205418} {\bibfield
  {journal} {\bibinfo  {journal} {Physical Review B}\ }\textbf {\bibinfo
  {volume} {75}},\ \bibinfo {pages} {205418} (\bibinfo {year}
  {2007})}\BibitemShut {NoStop}%
\bibitem [{\citenamefont {Kornyshev}\ and\ \citenamefont
  {Vorotyntsev}(1980)}]{kornyshev_nonlocal_1980}%
  \BibitemOpen
  \bibfield  {author} {\bibinfo {author} {\bibfnamefont {A.~A.}\ \bibnamefont
  {Kornyshev}}\ and\ \bibinfo {author} {\bibfnamefont {M.~A.}\ \bibnamefont
  {Vorotyntsev}},\ }\href
  {https://www.sciencedirect.com/science/article/abs/pii/003960288090597X}
  {\bibfield  {journal} {\bibinfo  {journal} {Surface Science}\ }\textbf
  {\bibinfo {volume} {101}},\ \bibinfo {pages} {23} (\bibinfo {year}
  {1980})}\BibitemShut {NoStop}%
\bibitem [{\citenamefont {Misra}\ and\ \citenamefont
  {Blankschtein}(2021)}]{misra_ion_2021}%
  \BibitemOpen
  \bibfield  {author} {\bibinfo {author} {\bibfnamefont {R.~P.}\ \bibnamefont
  {Misra}}\ and\ \bibinfo {author} {\bibfnamefont {D.}~\bibnamefont
  {Blankschtein}},\ }\href {\doibase 10.1021/acs.jpcc.0c09855} {\bibfield
  {journal} {\bibinfo  {journal} {The Journal of Physical Chemistry C}\
  }\textbf {\bibinfo {volume} {125}},\ \bibinfo {pages} {2666} (\bibinfo {year}
  {2021})}\BibitemShut {NoStop}%
\end{thebibliography}%
%

\end{document}

\title{Coupled interactions at the ionic graphene/water interface}
\author{Anton Robert, Hélène Berthoumieux and Marie-Laure Bocquet}

\maketitle
\tableofcontents{}

\subsubsection*{Notations}

Throughout those notes, the Fourier transform is defined as follows
:

\[
f(\mathbf{k})=\int\text{d}^{3}\mathbf{x}f(\mathbf{x})e^{-i\mathbf{k}\mathbf{x}}\ \ \ \ f(\mathbf{x})=\int\frac{\text{d}^{3}\mathbf{k}}{(2\pi)^{3}}f(\mathbf{k})e^{i\mathbf{k}\mathbf{x}}
\]
We use cylindrical coordinates in real $\mathbf{x}=(\mathbf{r},z)$
and reciprocal space $\mathbf{k}=(\mathbf{q},q_{z})$.

\section{Building dielectric response functions}

\subsection{Response functions}

We consider a thermodynamically closed system such that the term external
refers to something that is not a part of the physical system under
scrutiny. Under the application of a perturbating external electrostatic
potential $\phi_{\text{ext}}(\mathbf{x},t)$, it responds by generating
a charge density deviance $n_{\text{ind}}(\mathbf{x},t)$ - therefore
induced by the latter. We will focus on the statistically averaged
- denoted $\langle.\rangle$ - deviation produced by the system, in
space-time - denoted for short by $1\equiv(\mathbf{x}_{1},t_{1})$.
Note that Linear response theory relates both quantities via the response
function\emph{ -} or susceptibility \emph{- }$\chi$ as follows

\begin{equation}
\langle n_{\text{ind}}(1)\rangle=\int\text{d}2\chi(1,2)\phi_{\text{ext}}(2).\label{eq:from_linear_response}
\end{equation}
The total charge density of the system $n=n_{0}+n_{\text{ind}}$ might
not be equal to $n_{\text{ind}}$ if, without the external perturbation,
an inhomogeneity is already present in the system. In turn, the averaged
charge density deviance creates an averaged induced electrostatic
potential $\phi_{\text{ind}}$ that can be written

\begin{equation}
\langle\phi_{\text{ind}}(1)\rangle=\int\text{d}2v(1,2)\langle n_{\text{ind}}(2)\rangle.\label{eq:greens_function_method_induced_pot}
\end{equation}
Here we have used the \emph{Green's function method} \citep{schwinger_chapter_1998}
to solve Poisson's equation so that the kernel is the Coulomb interaction
between two particles of elementary charge (we use $e=1$) in vacuum
that reads
\begin{equation}
v(1,2)=\frac{1}{4\pi\epsilon_{0}}\frac{\delta(t_{1}-t_{2})}{\vert\mathbf{x}_{1}-\mathbf{x}_{2}\vert},\label{eq:coulomb_potential}
\end{equation}
where $\epsilon_{0}$ is the dielectric permittivity of vacuum. Accordingly,
we write Eq. \ref{eq:greens_function_method_induced_pot} for the
external potential $\phi_{\text{ext}}$ that arises from an external
charge density $n_{\text{ext}}$. The statistically averaged total
potential in the system is given by $\langle\phi_{\text{tot}}\rangle=\langle\phi_{\text{ext}}\rangle+\langle\phi_{\text{ind}}\rangle$.
Combining Eq. \ref{eq:from_linear_response} and Eq. \ref{eq:greens_function_method_induced_pot}
we can write the Green's function of the system $w(1,2)$ as 

\begin{equation}
w(1,1')=v(1,1')+\iint\text{d}2\text{d}3v(1,2)\chi(2,3)v(3,1').\label{eq:system_greens_function}
\end{equation}
The linearity of Poisson's equation makes the introduction of $w$
helpful because the total potential then reads

\begin{equation}
\langle\phi_{\text{tot}}(1)\rangle=\int\text{d}1'w(1,1')n_{\text{ext}}(1').\label{eq:total_potential}
\end{equation}
Those last equations directly establish the link between the averaged
total potential $\langle\phi_{\text{tot}}\rangle$ in the system and
$\chi$, its two-point susceptibility. Regarding the general structure
of $w$ in Eq. \ref{eq:system_greens_function}, it is constituted
of two objects linked by convolutions. To prepare future complexifications,
we represent by a diagram the bare potential $v$ and $\chi$. \begin{equation}\includegraphics{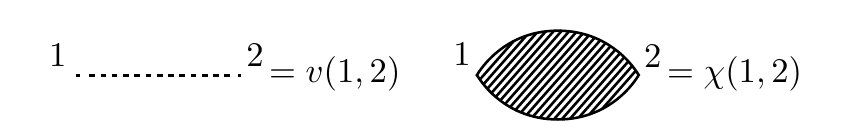}\label{fig:def_v_and_chi}\end{equation}We
define that their link represents a convolution in space-time. Therefore,
we can represent the Green's function as follows \begin{equation}\includegraphics{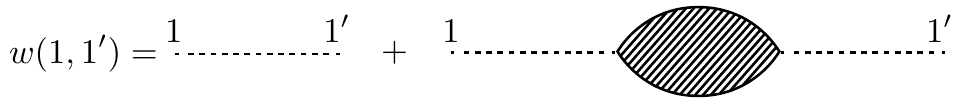}.\label{fig:greens_function}\end{equation}.

\subsection{Mean field correction: From $\chi^{(0)}$ to $\chi$.}

To build the response function $\chi$ of a system containing indistinguishable
particles, we start from the simplest possible version and enrich
the description from it. The first - drastic - physical assumption
is to consider that the particles are \emph{independent} so that we
first construct the \emph{non-interacting} response function $\chi^{(0)}$
using a microscopic model.

Enriching the description of the system requires considering interparticule
interactions. In this work we use a mean field method. Be it for electrons
in a metal \citep{mahan_many-particle_1990} or in simple liquids
\citep{hansen_theory_2013}, the mean field method is a well-known
renormalization scheme for homogeneous media to take into account
the collective behavior of particles. It amounts to considering that
independent particles respond to the external potential $\phi_{\text{ext}}$
plus the \emph{mean} polarization potential $\langle\phi_{\text{pol}}\rangle=v_{\text{inter}}*\langle n_{\text{ind}}\rangle$
of the other similar particles - here $v_{\text{inter}}$ is the interparticle
potential and $*$ denotes the space-time convolution. In other words,
the mean induced charge density is given by the sum of the two contributions
that reads
\begin{equation}
\langle n_{\text{ind}}(1)\rangle=\int\text{d}2\chi^{(0)}(12)\left[\phi_{\text{ext}}(2)+\int\text{d}3v_{\text{inter}}(23)\langle n_{\text{ind}}(3)\rangle\right].\label{eq:n_ind_mean_field}
\end{equation}
The recursive nature of this interparticle\emph{ renormalization scheme
}is captured by the integral equation for the response function $\chi$
which reads \begin{equation}\includegraphics{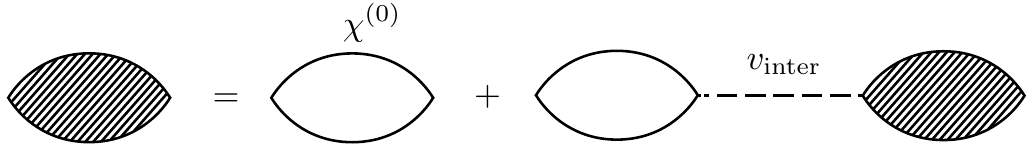},\label{fig:mean-field_chi}\end{equation}
or in equation

\begin{equation}
\chi(11')=\chi^{(0)}(11')+\iint\text{d}2\text{d}3\chi^{(0)}(12)v_{\text{inter}}(23)\chi(31').\label{eq:mean_field_equation}
\end{equation}
Note that the introduction of $\chi$ is a short-cut notation for
an infinite number of convolutions between $\chi^{(0)}$ and $v_{\text{inter}}$,
that are the elementary bricks. Indeed, by writing the beginning of
the infinite sum

\begin{equation}\includegraphics{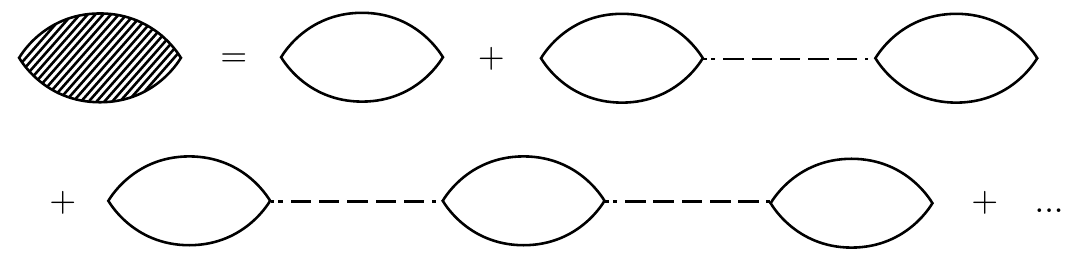},\label{fig:detail_chi_with_chi0}\end{equation}
we can observe that we have actually re-organized or \emph{renormalized}
an infinite sum of diagrams involving only $\chi^{(0)}$ and $v_{\text{inter}}$.
Turning to the topographic structure of $w$ when the diagram depicted
by $\chi$ is decomposed, it consists in enumerating all diagrams
that can be build with $\chi^{(0)}$ and $v_{\text{inter}}$, that
start and end with a Coulomb leg.

A special important case of mean-field renormalization is the random
phase approximation (RPA) \citep{bohm_collective_1953}. It is equivalent
to identifying the interparticle mean-field potential with the direct
potential - e.g. $v_{\text{inter}}=v$ for classical point charges.
For electrons, in the RPA, the potential can also include an exchange-correlation
potential, but those refinements are left out of this study. We attribute
the \emph{bare bubble} diagram to $\chi_{\text{e}}^{(0)}$ and the
\emph{hatched} \emph{bubble} diagram to the RPA response function
for electrons that we denote $\chi_{\text{e}}$ that reads \begin{equation}\includegraphics{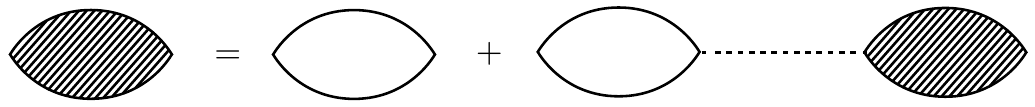}.\label{fig:detail_chi_with_chi0}\end{equation}

\section{The metal-like/liquid interface}

\begin{figure}[h]
\begin{centering}
\includegraphics[scale=0.7]{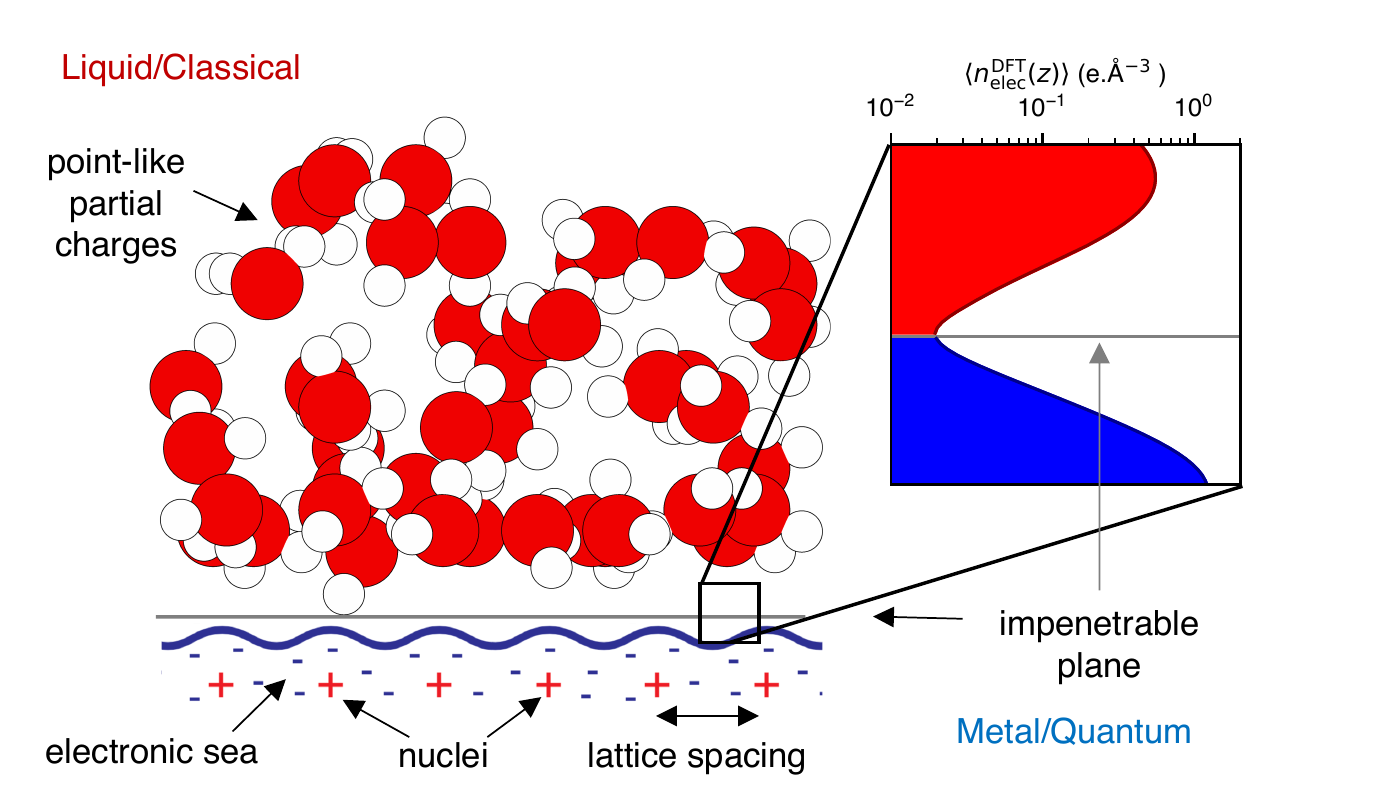}
\par\end{centering}
\caption{Schematic drawing of the metal/liquid interface. The zoom shows the
behavior of the total electronic density near the interface between
graphene and water. It is obtained with a static DFT calculation on
a \emph{ab initio} snapshot. \label{fig:Impenetrable-plane-at-1}}
\end{figure}

\subsection{Hamiltonian of the system}

In the framework of quantum field theory, we will use a bosonic operator
to depict the charge density of each medium $\hat{n}(\mathbf{x},t)$,
with $\mathbf{x}=(\mathbf{r},z)$ and $n=\langle\hat{n}\rangle$ is
the corresponding expectation value. It can either depicts the fictitious
charge density operator of the classical Gaussian liquid $\hat{n}_{\text{w}}$
\citet{kavokine_fluctuation-induced_2022} or the nuclear charge density
operator of the solid lattice $\hat{n}_{\text{n}}$. The creation
and annihilation Fermi fields that describe the behavior of the electrons
will be denoted $\Psi^{\dagger}(\mathbf{x},t)$ and $\Psi(\mathbf{x},t)$
such that the electronic charge density can be written $\hat{n}_{\text{e}}(\mathbf{x},t)=\Psi^{\dagger}(\mathbf{x},t)\Psi(\mathbf{x},t)$.
We consider the Coulomb interaction between partial atomic charges
on classical molecules in the liquid, electrons and nuclei in the
solid. The Hamiltonian of the system can be written $H=H_{0}+H_{\text{int}}$
with the quadratic Hamiltonian $H_{0}=H_{\text{w}}^{(0)}+H_{\text{e}}^{(0)}+H_{\text{n}}^{(0)},$ for
the electrons, liquid molecules and nuclei respectively that contains
one-particle terms only.

The crystal lattice, made of positively charged nuclei cannot be stable
without considering an implicit interaction with the electrons that
makes forces between nuclei nearly harmonic and not purely repulsive.
In turn, the eigenstates of the electrons differ from the ones of
the free electron gas because they feel the potential produced by
the clamp nuclei. Herein, we will be interested in time-averaged quantities
so that nuclei, with their core electrons, can be considered to be
fixed on their equilibrium positions. As a result, they do not participate
in the polarization charge and can be removed from the description
by considering an appropriate tight-binding model that takes into
account the lattice for valence electrons so that $H_{0}=H_{\text{w}}^{(0)}+H_{\text{e}}^{(\text{n})}$.
Ignoring exchange and correlation interactions between electrons,
we read

\begin{equation}
H_{\text{int}}=U_{\text{ee}}+U_{\text{ew}}+U_{\text{ww}},\label{eq:H_int}
\end{equation}
where the electronic interactions $U_{\text{ee}}$ have to be explicited
with the Fermi fields to remove the diverging self-energy term - i.e.

\begin{equation}
U_{\text{ee}}=\frac{1}{2}\iint\text{d}1\text{d}2\Psi^{\dagger}(1)\Psi^{\dagger}(2)v(1,2)\Psi(1)\Psi(2),\label{eq:elec_elec_interactions}
\end{equation}
but where the other interactions can be clearly expressed in a more
intuitive two-body form:

\begin{equation}
U_{\text{ew}}=\iint\text{d}1\text{d}2\hat{n}_{\text{e}}(1)v(1,2)\hat{n}_{\text{w}}(2),\label{eq:electron_water_interactions}
\end{equation}

\begin{equation}
U_{\text{ww}}=\frac{1}{2}\iint\text{d}1\text{d}2\hat{n}_{\text{w}}(1)v_{\text{w}}^{\text{eff}}(12)\hat{n}_{\text{w}}(2)\label{eq:water_water_interaction}
\end{equation}

\paragraph*{Liquid part}

For the liquid, the expression of $v_{\text{w}}^{\text{eff}}$ is
unknown. Nevertheless, by switching off the charge of electrons, we
can isolate the excess term $U_{\text{ww }}$as follows: 
\[
U_{\text{ww}}=H-H_{\text{w}}^{(0)}.
\]
The remaining Gaussian classical field is defined by its correlation
function so that those Hamiltonians can be written in the form of

\begin{equation}
H\left[n_{\text{w}}\right]=-\iint\text{d}1\text{d}2n_{\text{w}}(1)\chi_{\text{w}}^{-1}(12)n_{\text{w}}(2)\label{eq:gaussian_hamiltoniain-1}
\end{equation}

\begin{equation}
H_{\text{w}}^{(0)}\left[n_{\text{w}}\right]=-\iint\text{d}1\text{d}2n_{\text{w}}(1)\left[\chi_{\text{w}}^{(0)}\right]^{-1}(12)n_{\text{w}}(2)\label{eq:gaussian_hamiltonian_water_zero}
\end{equation}
Note that $\chi_{\text{w}}$ is the response function of \emph{the
liquid slab alone} and $\chi_{\text{w}}^{(0)}$ is the \emph{ideal}
or \emph{non-interacting} response function of the liquid slab. Consequently,
we can write a general expression for $v_{\text{w}}^{\text{eff}}$
using Eq. \ref{eq:water_water_interaction}, \ref{eq:gaussian_hamiltoniain-1},
\ref{eq:gaussian_hamiltonian_water_zero} that reads

\begin{equation}
v_{\text{w}}^{\text{eff}}(12)=\left[\chi_{\text{w}}^{(0)}\right]^{-1}(12)-\chi_{\text{w}}^{-1}(12).\label{eq:v_w_eff}
\end{equation}
Inserting Eq.\ref{eq:v_w_eff} in the mean field renormalization equation
Eq.\ref{eq:mean_field_equation}, we understand that it is satisfied
for $v_{\text{inter}}=v_{\text{w}}^{\text{eff}}$ so that $v_{\text{w}}^{\text{eff}}$
is the \emph{charge-charge mean field potential} \emph{in water}.
This prompts us to introduce a diagram for the response function of
the liquid. We have used the precedent diagrams for electrons and
we therefore introduce a new white and hatched ball-and-stick representation
for $\chi_{\text{w}}^{(0)}$ and $\chi_{\text{w}}$ respectively.
The mean field equation Eq. \ref{fig:mean-field_chi}, for the case
of a water slab alone reads
\begin{center}
\begin{equation}  \includegraphics{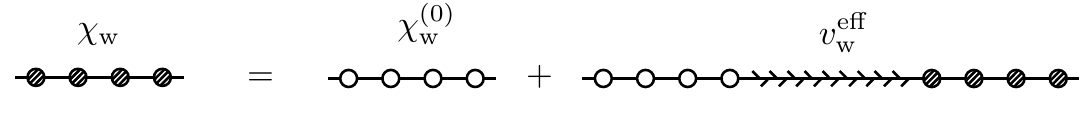} \label{fig:water_chi}  \end{equation}
\par\end{center}

\subsection{Response function at the interface \label{part:Many-body-Green's-function}}

From the previous sections, we have guessed some rules for constructing
diagrams that would lead to the Green's function of the system:
\begin{itemize}
\item \emph{(i)} Draw all possible \emph{linked} diagrams with $\chi_{\text{w}}^{(0)}$
and $\chi_{\text{e}}^{(0)}$ that start and ends with a Coulomb leg
$v$. The links are given by the mean-field potential between particles.
\item \emph{(ii)} \emph{Re-sum} the diagrams to make \emph{renormalized}
response functions appear.
\item \emph{(iii) }Write down the equation by reading the diagrams: a link
represents a \emph{convolution} in space-time.
\end{itemize}
Those rules stem from Feynman ones and are adapted to our case in
which we focus on response functions. They lead to a result that can
be derived with the functional formulation of quantum field theory
\citep{hedin_effects_1970,giustino_electron-phonon_2017}.

We re-organize all possible linked diagrams as prescribed by the aforementioned
rules. There are $2$ ways of re-summing the diagrams to put forward
either electrons or water molecules. We chose to separate the contribution
of the liquid alone. The Green's function of the liquid slab alone
reads \begin{equation}  \includegraphics{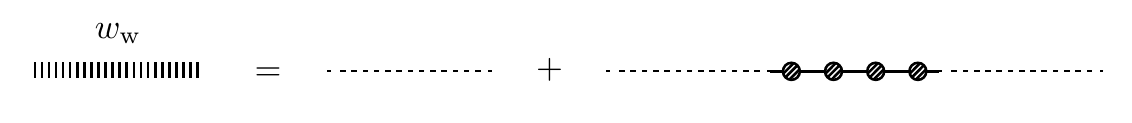}.\label{fig:w_w}  \end{equation}
The response function of electrons $\chi_{\text{e}}^{(0)}$ can now
be renormalized at the mean field level that we have seen in Eq. \ref{eq:mean_field_equation},
except that electrons interact via the Coulomb potential screened
by water so that $v_{\text{inter}}=w_{\text{w}}$. This means that
we need to introduce the \emph{in situ} response function of electrons
$\chi_{\text{e}}^{\text{(w)}}$ that reads \begin{equation}  \includegraphics{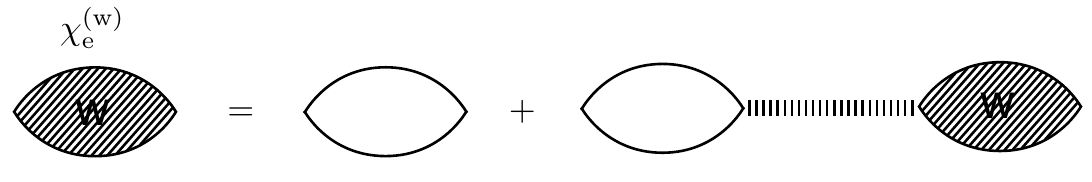}.\label{fig:chi_e^w}  \end{equation}
Finally, all diagrams in the Green's function of the system are contained
by expressing $w$ as follows \begin{equation}  \includegraphics{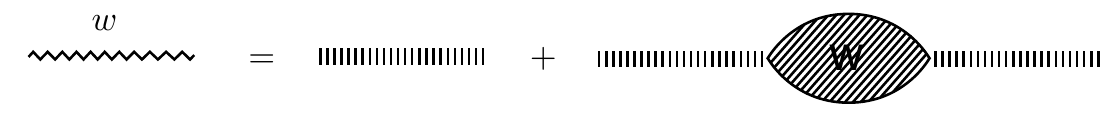}.\label{fig:w}  \end{equation}
The corresponding mathematical equations for Eq. \ref{fig:w_w},
\ref{fig:chi_e^w} and \ref{fig:w} read

\begin{align}
w_{\text{w}}(11')= & v(11')+\int\text{d}2\text{d}3v(12)\chi_{\text{w}}(23)v(31')\\
\chi_{\text{e}}^{\text{(w)}}(11')= & \chi_{\text{e}}^{(0)}(11')+\iint\text{d}2\text{d}3\chi_{\text{e}}^{(0)}(12)w_{\text{w}}(23)\chi_{\text{e}}^{\text{(w)}}(31')\label{eq:w_w}\\
w(11')= & w_{\text{w}}(11')+\iint\text{d}2\text{d}3w_{\text{w}}(12)\chi_{\text{e}}^{(\text{w})}(23)w_{\text{w}}(31')\label{eq:chi_e_w}
\end{align}

\subsection{Special case of the slit geometry}

Until now, our work does not depend on the geometry. If the system
has one interface, then Eq. \ref{fig:w} can be written in terms
of the well-identified response function $\chi_{\text{e}}^{\text{(w)}}$.
In contrast, when two interfaces are present, $\chi_{\text{e}}^{\text{(w)}}$
is the response function for the whole electronic part, that is the
two (semi-)metals with a channel filled with water in between. Due
to our finite matrix inversion for the liquid part, the finite size
channel geometry is necessary and we need to express $\chi_{\text{e}}^{\text{(w)}}$
with the response function of a single graphene sheet $\chi_{\text{e}}^{(0)}$.

Naively, we would like to split the electronic response into two parts,
one for the electrons at the bottom $\downarrow$ of the channel and
one for the electrons at the top $\uparrow$ that is \begin{equation}  \includegraphics{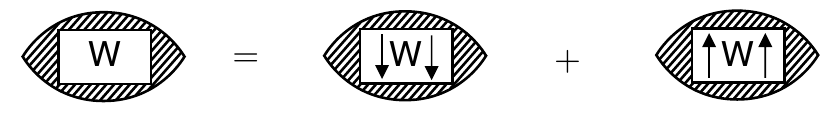},
\label{fig:chi_e^(w)}  \end{equation} where the diagrams on the
right-hand side are the renormalized response functions for the graphene
sheets at $z=0$ (bottom, $\downarrow$) and at $z=L$ (top, $\uparrow$).
The diagram on the left-hand side is therefore the first guess. The
correct result is not that simple because there are cross correlation
effects that have to be considered. Indeed, according to the aforementioned
rules, we miss some diagrams in $w$ and we need to combine top and
bottom response functions in all possible ways. We can separate all
combination in four categories depending on their first and last diagram
($\uparrow\uparrow$,$\uparrow\downarrow$, $\downarrow\uparrow$,
$\downarrow\downarrow$). The task is easier if we introduce the cross
correlation response function \begin{equation}  \includegraphics{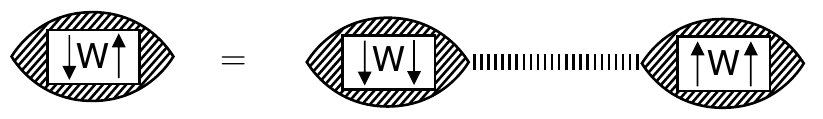},
\label{fig:chi_cross}  \end{equation} and its mean-field renormalized
analogue \begin{equation}  \includegraphics{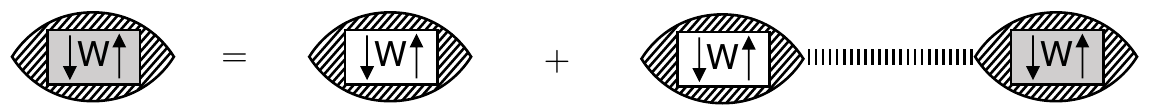}\label{fig:chi_cross_mean_field}  \end{equation}.
This gives the first two independent Dyson equations (Eq. \ref{fig:chi_cross_mean_field} 
can also be written for $\uparrow\downarrow$) and the remaining
two read \begin{equation}  \includegraphics{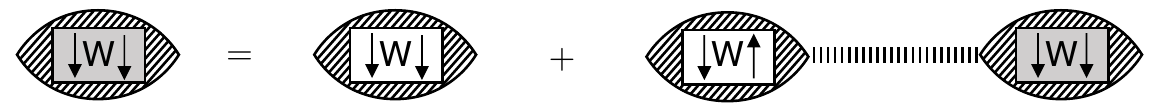},
\label{fig:chi_top_mean_field}  \end{equation} -- and the same
for $\uparrow\uparrow$. Assembling the four categories gives the
response function that we look for and

\begin{equation}  \includegraphics{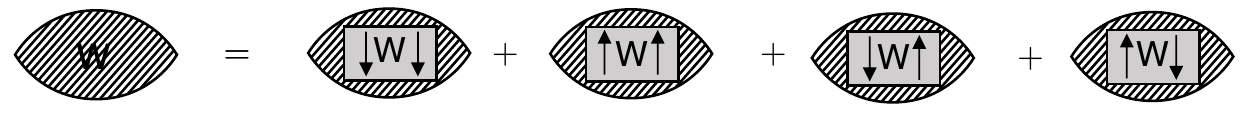}.
\label{fig:chi_e^(w)_slit}  \end{equation} It can be checked that
if the height of the channel tends to infinity, the cross susceptibilities
tend to $0$ and our initial assumption is correct - i.e. Eq. \ref{fig:chi_e^(w)_slit}
and Eq. \ref{fig:chi_e^(w)} are equal.

\section{Water model}

In this part, we build a versatile model for bulk water that can be
adapted to the slab geometry. It relies on the construction of the
non-interacting response function $\chi_{\text{w}}^{(0)}$ and the
effective potential between molecules $v_{\text{w}}^{\text{eff}}$.
We scrutinize the bulk medium before tackling the interface. Then
we express the local susceptibility $\bar{\chi}_{\text{w}}(z)$ in
our microscopic description.

\begin{figure}[h]
\begin{centering}
\includegraphics{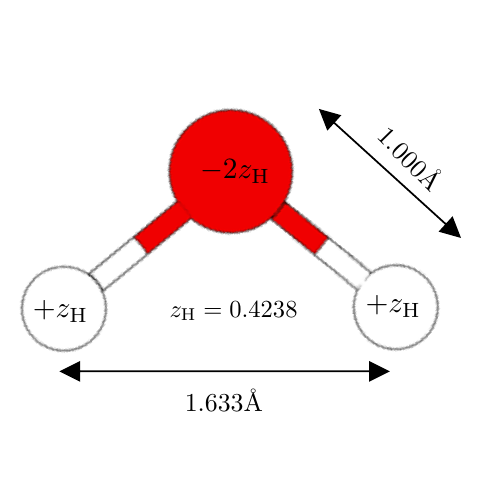}
\par\end{centering}
\caption{Sketch of a three-point charge SPC/E water molecule. Distances and partial charges are indicated on the figure.\label{fig:SPC/E-water-molecule.}}
\end{figure}

\subsection{Bulk water}

\subsubsection{Non-interacting susceptibility $\chi_{\text{w}}^{\text{(0)}}(k)$}

In this paragraph we detail our model for bulk water. First, we precise
that we will consider static external perturbations only. The time
integration in Eq.\ref{eq:from_linear_response} gives that the mean
induced charge density does not depend on time (as expected) and that
the relevant response quantity is the zero-frequency component of
the response function - e.g. $\chi_{\text{w}}(\mathbf{k})=\chi_{\text{w}}(\mathbf{k},\omega=0)$
in the homogeneous and isotropic system. We briefly recall how to
obtain it using Kramers-Kronig relationships and the fluctuation-dissipation
theorem \citet{kubo_fluctuation-dissipation_1966} :

\begin{align}
\chi_{\text{w}}(\mathbf{k}) & =\text{Re}\chi_{\text{w}}(\mathbf{k},\omega=0)\label{eq:static_bulk_response-1}\\
 & \overset{\text{KK}}{=}\int_{-\infty}^{+\infty}\frac{\text{d}\omega'}{\pi}\frac{\text{Im}\chi_{\text{w}}(\mathbf{k},\omega')}{\omega'}\\
 & \overset{\text{fluc-diss}}{=}\int_{-\infty}^{+\infty}\frac{\text{d}\omega'}{\pi}\frac{-\frac{\omega'}{2k_{B}T}S_{\text{w}}(\mathbf{k},\omega')}{\omega'}\\
 & =-\beta S_{\text{w}}(\mathbf{k},t=0)=-\beta S_{\text{w}}(\mathbf{k})\\
 & =-\frac{\beta}{\mathcal{V}}\left\langle n_{\text{w}}(\mathbf{k})n_{\text{w}}(-\mathbf{k})\right\rangle \label{eq:def_strucutre_factor}
\end{align}
where $\left\langle .\right\rangle $ means the phase-space averaged
at equilibrium, $S_{\text{w}}(1,2)=\left\langle n_{\text{w}}(1)n_{\text{w}}(2)\right\rangle $
is the charge structure factor, $\mathcal{V}$ the volume of the system.
The charge density of the water molecules $n_{\text{w}}(\mathbf{x})$
-- that must be replaced by $\delta n_{\text{w}}(\mathbf{x})=n_{\text{w}}(\mathbf{x})-\langle n_{\text{w}}(\mathbf{x})\rangle$
if the mean is not zero -- is composed of point-like charges. In
a field formulation, we can write the charge density as a convolution
between the fixed charge density $\sigma(\mathbf{x},\Omega)$ of one
rigid water molecule oriented with an angle $\Omega$, and the molecular
number density $N_{\text{w}}(\mathbf{x},\Omega)$ -- i.e. $\int_{\mathcal{V}}\int_{\Omega}N_{\text{w}}=N=n_{0}\mathcal{V}$
so that $n_{0}$ is the molecular density.

\begin{equation}
N_{\text{w}}(\mathbf{x},\Omega)=\sum_{i}\delta(\mathbf{x}-\mathbf{x}_{i})\delta(\Omega-\Omega_{i})\ \ \ \ \ \sigma(\mathbf{x},\Omega)=\sum_{\alpha}c_{\alpha}\delta(\mathbf{x}-\mathbf{s}_{\alpha}(\Omega)),\label{eq:number_density_sigma}
\end{equation}
They both depend on the molecule's orientation $\Omega$ that we do
not need to explicit here, but details can be found in e.g. \citet{jeanmairet_molecular_2013}.
The index $i$ runs on different molecules and $\alpha$ on different
atoms in the molecule with the partial charges $c_{\alpha}$. $\mathbf{s}_{\alpha}(\Omega)$
is the position of the atom $\alpha$ given the orientation
$\Omega$ of the molecule. Those quantities for the SPC/E model \citet{berendsen_missing_1987}
can be found in Fig. \ref{fig:SPC/E-water-molecule.}. This gives

\begin{equation}
n_{\text{w}}(\mathbf{x})=\int\text{d}\Omega\int\text{d}\mathbf{x}'\sigma(\mathbf{x}-\mathbf{x}',\Omega)N_{\text{w}}(\mathbf{x'},\Omega)\label{eq:charge_density_field}
\end{equation}
With those definitions, the structure factor reads

\begin{align}
S_{\text{w}}(\mathbf{k}) & =\frac{1}{\mathcal{V}}\iint\text{d}\Omega_{1}\text{d}\Omega_{2}\sigma(\mathbf{k},\Omega_{1})\sigma(-\mathbf{k},\Omega_{2})\left\langle N_{\text{w}}(\mathbf{k},\Omega_{1})N_{\text{w}}(-\mathbf{k},\Omega_{2})\right\rangle \label{eq:structure_factor}
\end{align}
The molecular number density field can be split into two parts
that highlight intramolecular and intermolecular correlations

\begin{equation}
\left\langle N_{\text{w}}(\mathbf{k},\Omega_{1})N_{\text{w}}(-\mathbf{k},\Omega_{2})\right\rangle =\langle\sum_{i}\delta(\Omega_{1}-\Omega_{i})\delta(\Omega_{2}-\Omega_{i})\rangle+\sum_{i,j\neq i}...\label{eq:number_number}
\end{equation}
Therefore, we can write $S_{\text{w}}(\mathbf{k})=S_{\text{w}}^{(0)}(\mathbf{k})+S_{\text{w}}^{(\text{inter)}}(\mathbf{k})$
with

\begin{align}
S_{\text{w}}^{(0)}(\mathbf{k}) & =\frac{1}{\mathcal{V}}\int\text{d}\Omega_{1}\sigma(\mathbf{k},\Omega_{1})\sigma(-\mathbf{k},\Omega_{1})\langle N_{\text{w}}(\Omega_{1})\rangle\label{eq:bare_strucutre_step1}
\end{align}
and the orientational density $N_{\text{w}}(\Omega_{1})=\sum_{i}\delta(\Omega-\Omega_{i})$.
In the bulk $\langle N_{\text{w}}(\Omega_{1})\rangle=1/8\pi^{2}$
is homogeneous and

\begin{equation}
S_{\text{w}}^{(0)}(\mathbf{k})=n_{0}\sum_{\alpha,\beta}c_{\alpha}c_{\beta}\int\frac{\text{d}\Omega_{1}}{8\pi^{2}}e^{-i\mathbf{k}(\mathbf{s}_{\alpha}(\Omega_{1})-\mathbf{s}_{\beta}(\Omega_{1}))}=n_{0}\sum_{\alpha,\beta}c_{\alpha}c_{\beta}j_{0}(kd_{\alpha\beta})\label{eq:orientation_average}
\end{equation}
where $j_{0}$ is the zeroth order spherical Bessel function, the
interatomic distances read $d_{\alpha\beta}=\vert\mathbf{s}_{\alpha}-\mathbf{s}_{\beta}\vert$
and $n_{0}$ is the molecular bulk density. For SPC/E water, we have

\begin{equation}
S_{\text{w}}^{(0)}(k)=n_{0}z_{\text{H}}^{2}\left[6-8\text{sinc}(kd_{\text{OH}})+2\text{sinc}(kd_{\text{HH}})\right]\label{eq:form_factor_SPCE_bulk}
\end{equation}
where $z_{\text{H}}$ is the partial charge on the hydrogen atom,
$d_{\text{OH}}$ and $d_{\text{HH}}$ are the bond distances of the
SPC/E molecule (see Fig. 2) \citet{berendsen_missing_1987} and $n_{0}=0.03298\ \text{Å}^{-3}$.
This gives the response function of the independent - or non-interacting
- molecules that reads
\begin{equation}
\chi_{\text{w}}^{\text{(0)}}(k)=-\beta S_{\text{w}}^{(0)}(k)\label{eq:non_interacting_bulk}
\end{equation}

\subsubsection{Effective potential $v_{\text{w}}^{\text{eff}}(k)$}

\begin{figure}[h]
\begin{centering}
\includegraphics{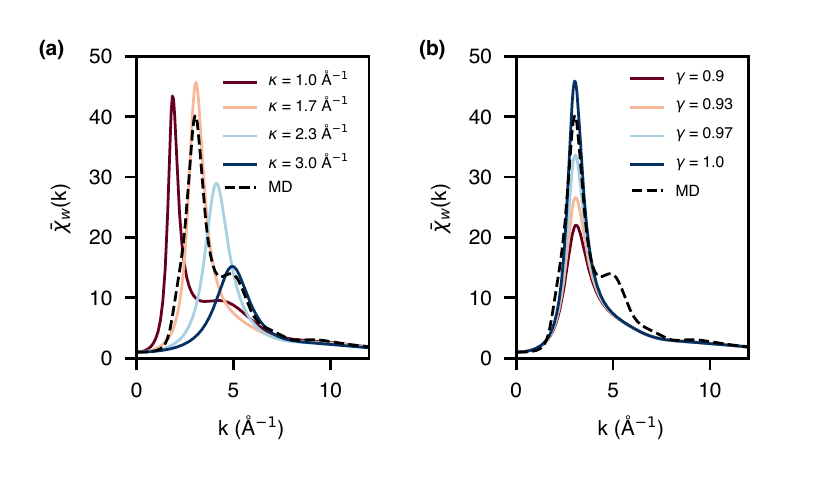}
\par\end{centering}
\caption{\HB{Sensitivity of $\bar{\chi}_{\text{w}}(k)$ to $\kappa$ and $\gamma$ variation. (a) Variation of $\kappa$ for $\gamma=0.99$. (b) Variation of $\gamma$ for $\kappa$=1.65~\AA$^{-1}$}
\label{fig:param_gamma_kappa}}

\end{figure}

In order to cast the water susceptibility into the mean-field equation
Eq. \ref{eq:mean_field_equation}, we want to infer the effective
intermolecular potential $v_{\text{w}}^{\text{eff}}(k)$ such that
the equation

\begin{equation}
\chi_{\text{w}}(k)=\chi_{\text{w}}^{\text{(0)}}(k)+\chi_{\text{w}}^{\text{(0)}}(k)v_{\text{w}}^{\text{eff}}(k)\chi_{\text{w}}(k)\label{eq:wRPA_fourier_bulk}
\end{equation}
is satisfied for the homogeneous, isotropic case. It leads to $v_{\text{w}}^{\text{eff}}(k)=1/\chi_{\text{w}}^{(0)}(k)-1/\chi_{\text{w}}(k)$
as already obtained in Eq. \ref{eq:v_w_eff}. The expression of $v_{\text{w}}^{\text{eff}}(k)$
is known in the long-wavelenght limit as we know that 
\begin{equation}
S_{\text{w}}^{(0)}(k)\xrightarrow[k\rightarrow0]{}n_{0}\frac{4}{3}k^{2}z_{\text{H}}^{2}d_{\text{OH}}^{2}\cos^{2}(\frac{\theta_{\text{HOH}}}{2})=n_{0}\frac{k^{2}\mu^{2}}{3}=k_{B}Tn_{0}\alpha k^{2}\label{eq:limit_S_0(k)}
\end{equation}
where we notice the Debye-Langevin polarizability $\alpha=\frac{\mu^{2}}{3k_{B}T}$
with $\mu=2.351$ D the dipole moment of the SPC/E water molecule.
Using the dimensionless susceptibility $\chi_{\text{w}}(k)=-\frac{4\pi\epsilon_{0}}{1}\frac{k^{2}}{4\pi}\bar{\chi}_{\text{w}}(k)$
with the limit $\bar{\chi}_{\text{w}}(k)\xrightarrow[k\rightarrow0]{}1-\frac{1}{\varepsilon_{\text{w}}}$,
we obtain

\begin{equation}
v_{\text{w}}^{\text{eff}}(k)\xrightarrow[k\rightarrow0]{}\frac{1}{4\pi\epsilon_{0}\varepsilon_{\text{w}}^{\text{eff}}}\frac{4\pi}{k^{2}}\label{eq:limit_veff}
\end{equation}
with the effective dielectric constant given by

\begin{equation}
\frac{1}{\varepsilon_{\text{w}}^{\text{eff}}}=\frac{1}{\bar{\chi}_{\text{w}}(k\rightarrow0)}-\frac{1}{\bar{\chi}_{\text{w}}^{(0)}(k\rightarrow0)}=1+\frac{1}{\varepsilon_{\text{w}}-1}-\frac{\epsilon_{0}}{n_{0}\alpha}.\label{eq:eps_eff}
\end{equation}
As expected, note that the intermolecular potential is zero --- or
$\varepsilon_{\text{w}}^{\text{eff}}\rightarrow\infty$ -- if water
molecules behave independently i.e. $\bar{\chi}_{\text{w}}\simeq\bar{\chi}_{\text{w}}^{(0)}$.
For water we have $\varepsilon_{\text{w}}^{\text{eff}}\simeq1.04$
so that the description of interacting molecular form factors with
a bare Coulomb potential is quite accurate in the long-wavelength
limit. For the remaining wavelengths, with the help of the results
from the bulk SPC/E MD \citet{jeanmairet_molecular_2016} shown in
the main text, we suggest the following ansatz (see main text for
formula in reciprocal space):

\begin{equation}
v_{\text{w}}^{\text{eff}}(x)=\frac{1-e^{-\kappa x}-x\gamma\frac{\kappa}{\pi}e^{-\kappa^{2}x^{2}/2}}{4\pi\epsilon_{0}\varepsilon_{\text{w}}^{\text{eff}}x}.\label{eq:v_eff_k}
\end{equation}
It gives a simple but accurate description of the response function.
An inverse screening length $\kappa$ dictates the position of the overscreening
peak and a prefactor $\gamma\simeq1$ modifies its amplitude. We have
found that $\kappa=1.65\text{Å}^{-1}$ and $\gamma=0.99$ can reproduce
the spectra of SPC/E water. The effect of those parameters on $\bar{\chi}_{\text{w}}(k)$
is shown in Fig. \ref{fig:param_gamma_kappa}. Note that the experimental
spectra \citet{bopp_static_1996} shows a less intensive peak that
can be easily fitted by tuning down $\gamma$. The second peak of
$\bar{\chi}_{\text{w}}(k)$ around $k\simeq5\text{\AA}^{-1}$can also
be included to refine the model, but we expect no important change
on the long-range collective dielectric response of water. 

\subsection{Interfacial water}

\subsubsection{Non-interacting susceptibility $\chi_{\text{w}}^{\text{(0)}}(q,z,z')$}

Looking for the response function $\chi_{\text{w}}(q,z,z')$ at interfaces,
we can proceed from Eq. \ref{eq:static_bulk_response-1} to Eq. \ref{eq:def_strucutre_factor}
to find a relation between the susceptibility and the corresponding
structure factor. Equally, we dwell on the non-interacting part and
find 
\begin{equation}
\chi_{\text{w}}^{(0)}(q,z,z')=-\beta S_{\text{w}}^{(0)}(q,z,z')\label{eq:chi_strucutre_0}
\end{equation}
We proceed like in the bulk and after some straightforward steps, we
obtain

\begin{equation}
S_{\text{w}}^{(0)}(\mathbf{q},z,z')=\frac{1}{\mathcal{A}}\iint\text{d}\Omega_{1}\text{d}z_{1}\sigma(\mathbf{q},z-z_{1},\Omega_{1})\sigma(-\mathbf{q},z'-z_{1},\Omega_{1})\langle N_{\text{w}}(z_{1},\Omega_{1})\rangle\label{eq:init_structure_factor_interface}
\end{equation}
where $\mathcal{A}$ is the surface area of the interface and

\begin{equation}
\sigma(\mathbf{q},z,\Omega)=\sum_{\alpha}c_{\alpha}e^{-i\mathbf{q}\mathbf{s}_{\alpha}(\Omega)}\delta(z-z_{\alpha}(\Omega))\ \ \ \ \ \ N_{\text{w}}(z_{1},\Omega)=\sum_{i}\delta(\Omega_{1}-\Omega_{i})\delta(z_{1}-z_{i})\label{eq:number_density_inter}
\end{equation}
Looking at the structure of the product $\sigma(\mathbf{q},z,\Omega_{1})\sigma(-\mathbf{q},z',\Omega_{1})$,
we can express with the sum and differences of $z$ and $z'$ as follows

\begin{align}
\sigma(\mathbf{q},z,\Omega_{1})\sigma(-\mathbf{q},z',\Omega_{1}) & =\sum_{\alpha\beta}c_{\alpha}c_{\beta}e^{-i\mathbf{q}(\mathbf{s}_{\alpha}(\Omega)-\mathbf{s}_{\beta}(\Omega))}\delta\left(z-z_{\alpha}(\Omega)\right)\delta\left(z'-z_{\beta}(\Omega)\right)\nonumber \\
 & =\sum_{\alpha\beta}c_{\alpha}c_{\beta}e^{-i\mathbf{q}(\mathbf{s}_{\alpha}(\Omega)-\mathbf{s}_{\beta}(\Omega))}\delta\left(z-z'-\left[z_{\alpha}(\Omega)-z_{\beta}(\Omega)\right]\right)\times\delta(z+z'-\left[z_{\alpha}(\Omega)+z_{\beta}(\Omega)\right])\label{eq:sum_and__diff_of_z_zp}
\end{align}
This makes the convolution of Eq. \ref{eq:init_structure_factor_interface}
with $N_{\text{w}}(z_{1},\Omega_{1})$ possible. It reads

\begin{equation}
S_{\text{w}}^{(0)}(\mathbf{q},z,z')=\frac{1}{\mathcal{A}}\int\text{d}\Omega\sum_{\alpha\beta}c_{\alpha}c_{\beta}e^{-i\mathbf{q}(\mathbf{s}_{\alpha}(\Omega)-\mathbf{s}_{\beta}(\Omega))}\delta\left(z-z'-\left[z_{\alpha}(\Omega)-z_{\beta}(\Omega)\right]\right)\times N_{\text{w}}\left(\frac{z+z'-\left[z_{\alpha}(\Omega)+z_{\beta}(\Omega)\right]}{2},\Omega\right)\label{eq:S(q)_with_Nw}
\end{equation}
For an homogeneous density, we would be able to get the term involving
$N_{\text{w}}$ out of the summation. Assuming equiprobable orientation
of molecules in the entire slab, we would recover the inverse Fourier
transform of the bulk structure factor, i.e.

\begin{align}
S_{\text{w}}^{(0)}(\mathbf{q},\vert z-z'\vert) & =\int\frac{\text{d}q_{z}}{2\pi}e^{iq_{z}\vert z-z'\vert}S_{\text{w}}^{(0)}(\mathbf{k})\label{eq:inverse_fourier_transform_S(q)}\\
 & =n_{0}z_{H}^{2}\left[6\delta(z-z')-8I(q,d_{\text{OH}},\vert z-z'\vert)+2I(q,d_{\text{HH}},\vert z-z'\vert)\right]\label{eq:analytical_value}
\end{align}
with $I(q,d,z)=J_{0}(q\sqrt{d^{2}-z^{2}})\Theta(d-\vert z\vert)/2d$
and $J_{0}$ is the zeroth order Bessel function. We work in this
direction and try to express $S_{\text{w}}^{(0)}(\mathbf{q},z,z')$
with $S_{\text{w}}^{(0)}(\mathbf{q},\vert z-z'\vert)$. Using the
condition enforced by the first Dirac delta function in Eq. \ref{eq:S(q)_with_Nw}
-- $z=z'+\left[z_{\alpha}(\Omega)-z_{\beta}(\Omega)\right]$ --
we can replace the term involving $N_{\text{w}}$ by either $N_{\text{w}}\left(z'-z_{\beta}(\Omega),\Omega\right)$,
or $N_{\text{w}}\left(z-z_{\alpha}(\Omega),\Omega\right)$, but also
by $\sqrt{N_{\text{w}}\left(z-z_{\alpha}(\Omega),\Omega\right)N_{\text{w}}\left(z'-z_{\beta}(\Omega),\Omega\right)}$,
without any approximation. Making now the approximation $N_{\text{w}}\left(z-z_{\alpha},\Omega\right)\simeq N_{\text{w}}\left(z,\Omega\right)$
and assuming equiprobable orientation of molecules in the entire slab
we can write

\begin{equation}
S_{\text{w}}^{(0)}(q,z,z')\simeq\frac{\sqrt{n_{0}(z)n_{0}(z')}}{n_{0}}S_{\text{w}}^{(0)}(q,\vert z-z'\vert),\label{eq:S_w_z_zp}
\end{equation}
which is the main result of this paragraph. This approximation is
valid if the molecular profile typically varies on a scale larger
than the size of a water molecule. Note however that the Taylor expansion
of $N_{\text{w}}$ near the interface give rises to terms linear in
$z_{\alpha}(\Omega)$ that can most probably gives a zero contribution
when the angular integration is carried out (under the approximation
of equiprobable orientation). Combining Eq. \ref{eq:chi_structure_0}
and Eq. \ref{eq:S_w_z_zp} gives the equation (Eq. 4) of the main text.

\subsubsection{Effective potential $v_{\text{w}}^{\text{eff}}(q,\vert z-z'\vert)$}

What is the charge-charge effective potential in the water slab ?
We don't know it, but we can suppose that \emph{water molecules interact
in the slab as if they were in the bulk. }This is a common approximation
in the liquid state theories of water - e.g.\citet{jeanmairet_molecular_2013}.
We therefore take $v_{\text{w}}^{\text{eff}}$ from Eq. \ref{eq:v_eff_k}
and give for completeness the appropriate Fourier transform that reads

\begin{equation}
v_{\text{w}}^{\text{eff}}(q,\vert z-z'\vert)=\frac{1}{4\pi\epsilon_{0}\varepsilon_{\text{w}}^{\text{eff}}}\left(\frac{2\pi}{q}e^{-q\vert z-z'\vert}-\frac{2\pi}{Q}e^{-Q\vert z-z'\vert}-\gamma\frac{2\pi}{\kappa}e^{-q^{2}/2\kappa^{2}}\frac{e^{-\kappa^{2}(z-z')^{2}}}{\pi}\right),\label{eq:v_eff_q}
\end{equation}
where $Q^{2}=q^{2}+\kappa^{2}$. Those expressions are all linked
by Fourier transform.

\subsubsection{Matrix filling}

The matrix $X_{\text{w}}^{(0)}$ has a size $\left\lfloor L/\text{d}z\right\rfloor \times\left\lfloor L/\text{d}z\right\rfloor $
with grid spacing $\text{d}z=0.02\text{\AA}$ and length $L=6\text{nm}$.
We rely on the condition that an homogeneous external potential cannot
induce a charge density disturbance (i.e. $\int\text{d}z'\chi_{\text{w}}^{(0)}(z,z')=0$)
to fill the matrix. This can be checked in the bulk and imposed at
altitudes close to the molecular density depletion. Therefore, in
order to obtain the entire matrix $X_{\text{w}}^{(0)}$ and avoid
numerical integration errors due to the Dirac delta functions, we
fill the non-diagonal entries of $X_{\text{w}}^{(0)}$ according to
$\chi_{\text{w}}^{(0)}(q,z,z')=-\beta S_{\text{w}}^{(0)}(q,z,z')$
and impose that all lines and column sum to 0 to fill the diagonal.

\section{Local dielectric susceptibility $\bar{\chi}_{\text{w}}(z)$}

\subsection{Expression from $\chi_{\text{w}}(q,z,z')$}

In order to confirm our model for water,  we \HB{define}  the \HB{following} non-local dielectric function

\begin{equation}
\varepsilon_{\text{w}}^{-1}(q,z,z')=\delta(z-z')+\int\text{d}z_{1}v(q,z,z_{1})\chi_{\text{w}}(q,z_{1},z')\label{eq:non-local_dielectric_funcntion}
\end{equation}
with the converged grid spacing $\text{d}z=0.02\text{Å}$. \HB{The expression for the kernel $\varepsilon_{\text{w}}^{-1}(r,r')$ is derived from the bulk relation between $\langle\phi_{\rm tot}\rangle$
	and $\phi_{\rm ext}$,
\begin{equation}
	\langle\phi_{\rm tot}\rangle(r)= \phi_{\rm ext}(r)+\int d^3r"d^3r'v(r,r")\chi_{\rm w}(r",r')\phi_{\rm ext}(r') 
\end{equation}
that we obtained using Eqs. (\ref{eq:system_greens_function},\ref{eq:total_potential})}

At this
point, we still have a function of three variables that we cannot
easily visualize. We therefore want to link $\varepsilon_{\text{w}}^{-1}(q,z,z')$
to the inverse local dielectric function $\varepsilon_{\text{w}}^{-1}(z)$
that relates the total electric field $E_{z}(z)$ in the system to
the constant one that is applied. Again, instead of turning back to
electrostatics, we construct microscopically the external electric
field. We place two infinite plates of opposite surface charge that
sandwich from very far the system under scrutiny. The external electric
field in the system is constant and equal to $D_{z}/\epsilon_{0}$.
Seen by the system of finite size under scrutiny, this homogeneous
(in the plane), constant displacement field applied in the direction
normal to the surface can be written as $\phi_{\text{ext}}(z)=-\frac{D_{z}}{\epsilon_{0}}z$
plus a constant that we set to zero. The electric field in the system
reads $\partial_{z}\phi_{\text{tot}}(z)=-E_{z}(z)$. Also, for the
microscopic linear response framework that we use, the electrostatic
potentials are linked as follows :

\begin{equation}
\phi_{\text{tot}}(z)=\int_{0}^{L}\varepsilon_{\text{w}}^{-1}(q\rightarrow0,z,z')\phi_{\text{ext}}(z')\text{d}z'\label{eq:starting_local}
\end{equation}
The electric field and the constant displacement field are linked
via the local dielectric function such that

\begin{equation}
E_{z}(z)=\epsilon_{0}^{-1}\varepsilon_{\text{w}}^{-1}(z)D_{z}\label{eq:electric_field_and_displacement}
\end{equation}
We can differentiate Eq. \ref{eq:starting_local} and use Eq. \ref{eq:electric_field_and_displacement}
to obtain

\begin{equation}
\varepsilon_{\text{w}}^{-1}(z)=\partial_{z}\int_{0}^{L}\text{d}z'\varepsilon_{\text{w}}^{-1}(q\rightarrow0,z,z')z'\label{eq:local_dielectric_functionn}
\end{equation}
from which the local susceptibility reads $\bar{\chi}_{\text{w}}(z)=1-\varepsilon_{\text{w}}^{-1}(z)$.
It relates the polarization of the medium to the applied electric
field $\epsilon_{0}E_{z}(z)=D_{z}-P_{z}(z)$ where $P_{z}(z)=\bar{\chi}(z)D_{z}$.

\subsection{Comparison with Landau-Ginzburg model}

\begin{figure}[h]
\begin{centering}
\includegraphics{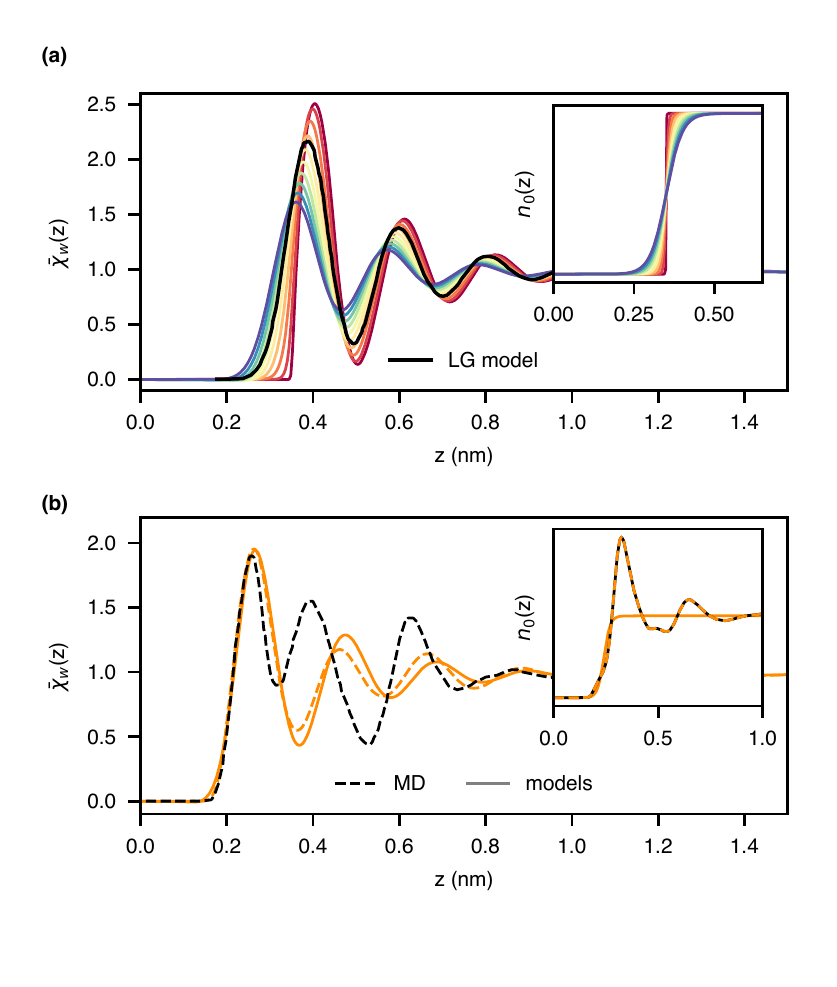}
\par\end{centering}
\caption{\textbf{(a)} Local dielectric susceptibility of the slab $P_{z}=\bar{\chi}_{\text{w}}(z)D_{z}$
computed from the non-local response function $\chi_{\text{w}}(q,z,z')$
with varying smoothness of the step function $\sigma$ and compared
with Landau-Ginzburg model of Ref. \citep{monet_nonlocal_2021}. \textbf{(b)}
Models of the main text compared with the results from the MD simulation
of \citep{monet_nonlocal_2021}. \MLB{The force field parameters chosen for water and graphene are taken from Ref. \citep{Werder2003}.} 
\label{fig:Local-dielectric-susceptibility}}
\end{figure}

The smoothed step function molecular density profile $n_{0}(z)$ is
chosen because:
\begin{itemize}
\item the use of $v_{\text{w}}^{\text{eff}}$ that implies that molecules
interact as is they were in the bulk, that is with an homogeneous
molecular density $n_{0}$.
\item it gives the same $\bar{\chi}_{\text{w}}(z)$ we obtain by using the
real hydrogen density from the MD simulation. We use the hydrogen atoms because
they can go at lower altitudes than the oxygen atoms.
\end{itemize}
Its expression is given by 
\[
n_{0}(z)=\frac{n_{0}}{4}\left[\tanh(\frac{z-d_{0}}{\sigma_{0}})+1\right]\left[\tanh(\frac{L-z-d_{0}}{\sigma_{0}})+1\right].
\]
We can see the effect of the smoothness in Fig. \ref{fig:Local-dielectric-susceptibility}a.
The agreement with the Landau-Ginzburg model introduced in \citep{monet_nonlocal_2021}
is excellent for $\sigma_{0}=0.2\text{Å}$. Compared to the MD results
\citep{monet_nonlocal_2021} in Fig. \ref{fig:Local-dielectric-susceptibility}b,
the use of $\sigma_{0}=0.3\text{Å}$ gives the correct amplitude for
the first peak. The oscillation period, amplitude and decay length
are also in good agreement. By tuning the molecular density profile,
we can fit exactly the MD profile, but our goal is not to reproduce
the local dielectric of an MD simulation that uses effective Lennard-Jones
potentials or effective interactions between the surface and the water
molecules -- graphene must be absent when we build $\chi_{\text{w}}$.

\section{Graphene model}

Here we give the expression for the non-interacting and the interacting response function
of graphene. 

\subsection{The non-interacting susceptibility $\chi_{\text{e}}^{(\text{0})}(q,z,z')$}

The non-interacting response function of graphene sheet $\chi_{\text{e}}^{(0)}(q)$
can be calculated with a tight--binding model at $T=0\text{K}$ \citet{hwang_dielectric_2007}.
It reads

\[
\chi_{\text{e}}^{(0)}(q)=-\frac{2k_{F}}{\hbar v_{F}\pi}\left[1+\frac{1}{2}\Theta\left(\frac{q}{2k_{F}}-1\right)\left(\frac{\pi}{2}\frac{q}{2k_{F}}-\sqrt{1-\left(\frac{2k_{F}}{q}\right)^{2}-\frac{q}{2k_{F}}}\arcsin\left(\frac{2k_{F}}{q}\right)\right)\right]
\]
where $v_{F}=1\ \text{nm}.\text{fs}^{-1}$ is the Fermi velocity and
$k_{F}=E_{F}/\hbar v_{F}$ is the Fermi wave vector and $E_{F}$ the
Fermi energy. The effect of temperature and Fermi level $E_{F}$ is
not investigated in this article and we rather use a minimal Fermi
level doping $E_{F}=k_{B}T$ that represents the minimal concentration
of mobile electrons.

\subsection{The interacting susceptibility $\chi_{\text{e}}^{(\text{w})}(q,z,z')$}

For two confining plates centered in $z=0$ and $z=L$, the expression
of the electronic response function is obtained by following the steps
from Eq. \ref{fig:chi_e^(w)} to Eq. \ref{fig:chi_e^(w)_slit}.
This gives

\begin{equation}
\chi_{\text{e}}^{(\text{w})}(q,z,z')=\chi_{\text{e}\uparrow}^{(\text{w})}(q)\delta(z)\delta(z')+\chi_{\text{e}\downarrow}^{(\text{w})}(q)\delta(L-z)\delta(L-z')+\chi_{\text{e}\uparrow\downarrow}^{(\text{w})}(q)\delta(z-L)\delta(z')+\chi_{\text{e}\downarrow\uparrow}^{(\text{w})}(q)\delta(z)\delta(L-z')\label{eq:chi_e_graphene_slits}
\end{equation}
with

\begin{equation}
\chi_{\text{e}\uparrow}^{(\text{w})}(q)=\frac{\chi_{\text{e}/\text{Gr}}^{(\text{w})}(q)}{1-\left[w_{\text{w}}(q,0,L)\chi_{\text{e}/\text{Gr}}^{(\text{w})}(q)\right]^{2}}\ \ \ \ \ \ \ \chi_{\text{e}\downarrow}^{(\text{w})}(q)=\chi_{\text{e}\uparrow}^{(\text{w})}(q)\label{eq:chi_up}
\end{equation}

\begin{equation}
\chi_{\text{e}\uparrow\downarrow}^{(\text{w})}(q)=\frac{\chi_{\text{e}/\text{Gr}}^{(\text{w})}(q)w_{\text{w}}(q,0,L)\chi_{\text{e}/\text{Gr}}^{(\text{w})}(q)}{1-\left[w_{\text{w}}(q,0,L)\chi_{\text{e}/\text{Gr}}^{(\text{w})}(q)\right]^{2}}\ \ \ \ \ \ \ \chi_{\text{e}\downarrow\uparrow}^{(\text{w})}(q)=\chi_{\text{e}\uparrow\downarrow}^{(\text{w})}(q)\label{eq:chi_cross}
\end{equation}
and where

\begin{equation}
\chi_{\text{e/Gr}}^{\text{(w)}}(q)=\frac{\chi_{\text{e}}^{(0)}(q)}{1-w_{\text{w}}(q,0,0)\chi_{\text{e}}^{(0)}(q)}\label{eq:chi_e_w_appendix_with_matrix}
\end{equation}
The ``uncoupled'' and ``semi-coupled'' curves in the main text
are obtained by replacing $w_{\text{w}}(q,0,0)$ with $v(q,0,0)=\frac{1}{4\pi\epsilon_{0}}\frac{2\pi}{q}$
in Eq. \ref{eq:chi_e_w_appendix_with_matrix}.

\section{Supplementary discussions}

\subsection{Long-wavelength limit error for water}

In the case of the infinite height channel $L\rightarrow\infty$,
Eq. \ref{eq:chi_e_graphene_slits} reduces to $\chi_{\text{e}}^{(\text{w})}(q,z,z')=\chi_{\text{e/Gr}}^{\text{(w)}}\delta(z)\delta(z')$
such that the system's Green's function reads
\begin{equation}
w(q,z,z')=w_{\text{w}}(q,z,z')+w_{\text{w}}(q,z,0)\chi_{\text{e}}^{(\text{w})}(q)w_{\text{w}}(q,0,z').\label{eq:case_infinite-1}
\end{equation}
Close to the surface, for $z\simeq z'\simeq0$, if we take the long-wavelength
limit according Eq. \ref{eq:limit_w_w}, the second on the rhs of
Eq. \ref{eq:case_infinite-1} is reduced by a factor a roughly $(\varepsilon_{\text{w}}^{*})^{2}\simeq1600$
for water, as first stated in \citet{kornyshev_nonlocal_1980}.

\subsection{\HB{Comparison with other studies considering a metallic behavior of the surface}}

\begin{figure}[h]
\begin{centering}
\includegraphics{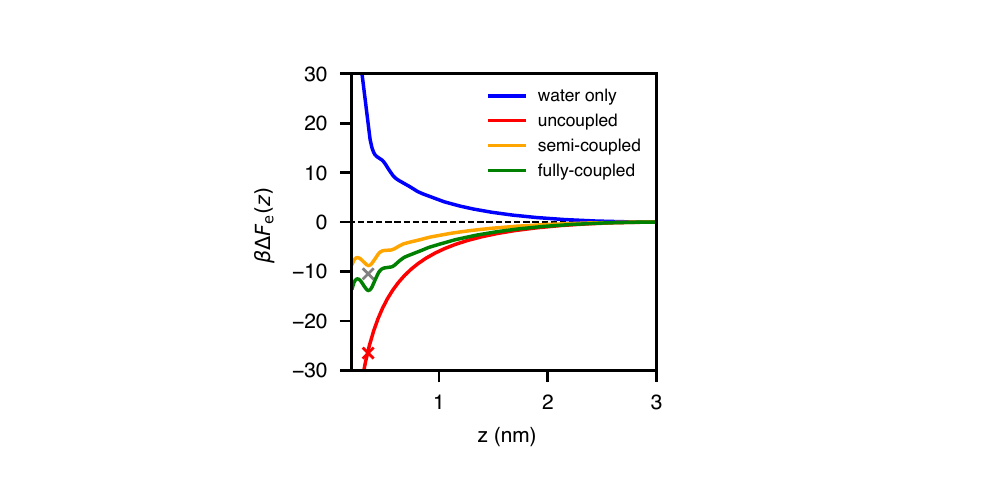}
\par\end{centering}
\caption{ Details of the contribution of the PMF shown in the main
text. The crosses are results extracted from Ref. \citep{misra_ion_2021}.
\label{fig:Details-of-the}}
\end{figure}

To evaluate the role of the graphene sheet, we can isolate the electronic
part $F_{\text{e}}=F-F_{\text{w}}$, where $F_{\text{w}}$ solely
contains the contribution of water and is obtained via the equation (Eq. 7)
of the main text, replacing $w$ by $w_{\text{w}}$. The details of
the contributions are shown in Fig. \ref{fig:Details-of-the}. \HB{Note that the blue curve 'water only' corresponds to $\Delta F$ as defined in Eq. (7) and is represented to permit an estimation of the electronic stabilization amplitude. In this case, the electronic contribution vanishes.} We
report on the same figure the results of Ref. \citep{misra_ion_2021}
that decomposes the wall-ion contribution for the adsorption site
of SCN$^{-}$, which is an ion too large and anisotropic to be quantitatively
compared with isotropic ion in presence of water. Nevertheless, the
good agreement with our results regarding metal-ion potential attenuation
validates our description of the system. 

\HB{Note that it is not possible to disentangle electronic and water contribution of the PMF when metallic character of the graphene surface is neglected, which is the case in classical MD simulations.  }
\subsection{Electron-electron potential attenuation}

For an infinite channel height $L\rightarrow\infty$, we have

\begin{equation}
w_{\text{w}}(q,0,0)=\frac{1}{4\pi\epsilon_{0}}\frac{2\pi}{q}\left[1-g_{\text{w}}(q)\right],\label{eq:effective_electron_potentials}
\end{equation}
where $g_{\text{w}}(q)$ is the surface response function of the water
medium, defined as

\begin{equation}
g_{\text{w}}(q)=-\frac{1}{4\pi\epsilon_{0}}\frac{2\pi}{q}\iint_{0}^{+\infty}\text{d}z\text{d}z'e^{-q(z+z')}\chi_{\text{w}}(q,z,z'),\label{eq:SRF}
\end{equation}
and that converges to the image charge coefficient in the long-wavelength
limit \citet{kavokine_fluctuation-induced_2022}

\begin{equation}
g_{\text{w}}(q)\xrightarrow[q\rightarrow0]{}\frac{\varepsilon_{\text{w}}-1}{\varepsilon_{\text{w}}+1}.\label{eq:limit_g}
\end{equation}
This gives a two-dimensional screening potential reduced by a factor
$\varepsilon_{\text{w}}^{*}=(\varepsilon_{\text{w}}+1)/2$ such that
\begin{equation}
w_{\text{w}}(q,0,0)\xrightarrow[q\rightarrow0]{}\frac{1}{4\pi\epsilon_{0}\varepsilon_{\text{w}}^{*}}\frac{2\pi}{q}.\label{eq:limit_w_w}
\end{equation}
This is the potential by which electrons interact in graphene.

\bibliographystyle{apsrev4-1}
%\bibliography{main}
%